\documentclass[12pt]{article}
\usepackage{amsmath,latexsym,epsf}
\usepackage{epsfig,amsfonts,cite}
\usepackage{amssymb}
\usepackage{graphicx}

\newcommand{\pos}[1]{\mathbf{#1}}

\begin{document}

\title{Finite-Size Scaling in 
the Driven Lattice Gas
}
\date{}
\author{
  \\
  { Sergio Caracciolo}              \\
  {\small\it Dipartimento di Fisica and INFN, Universit\`a di Milano, 
    and INFM-NEST, }  \\[-0.2cm]
  {\small\it I-20133 Milano, ITALY}          \\[-0.2cm]
  {\small Internet: {\tt sergio.caracciolo@mi.infn.it}}   
  \\[-0.1cm]  \and
  { Andrea Gambassi }              \\
  {\small\it Max-Planck-Institut f\"ur Metallforschung,}  \\[-0.2cm]
  {\small\it Heisenbergstr. 3, D-70569, Stuttgart, GERMANY,}      \\[-0.2cm]
  {\small\it and Institut f\"ur Theoretische und Angewandte Physik,} \\[-0.2cm]
  {\small\it Universit\"at Stuttgart, Pfaffenwaldring 57, 
       D-70569 Stuttgart, GERMANY} \\[-0.2cm]
  {\small Internet: {\tt gambassi@mf.mpg.de}}   
  \\[-0.1cm]  \and
  { Massimiliano Gubinelli}              \\
  {\small\it Dipartimento di Matematica Applicata  and INFN,
       Universit\`a di Pisa} \\[-0.2cm]
  {\small\it I-56100 Pisa, ITALY}          \\[-0.2cm]
  {\small Internet: {\tt m.gubinelli@dma.unipi.it}}   
  \\[-0.1cm]  \and
  { Andrea Pelissetto}              \\
  {\small\it Dipartimento di Fisica and INFN,
           Universit\`a di Roma ``La Sapienza''}        \\[-0.2cm]
  {\small\it I-00185 Roma, ITALY}          \\[-0.2cm]
  {\small Internet: {\tt Andrea.Pelissetto@roma1.infn.it}}   \\[-0.2cm]
  {\protect\makebox[5in]{\quad}}  
  \\
}
\vspace{0.5cm}

\newcommand{\be}{\begin{equation}}
\newcommand{\<}{\langle}
\renewcommand{\>}{\rangle}
\newcommand{\R}[2]{$(#2,#1)$}
\def\spose#1{\hbox to 0pt{#1\hss}}
\def\ltapprox{\mathrel{\spose{\lower 3pt\hbox{$\mathchar"218$}}
 \raise 2.0pt\hbox{$\mathchar"13C$}}}
\def\gtapprox{\mathrel{\spose{\lower 3pt\hbox{$\mathchar"218$}}
 \raise 2.0pt\hbox{$\mathchar"13E$}}}

\newcommand{\co}{ {\cal O}}

\maketitle
\thispagestyle{empty}   

\vspace{0.2cm}

\begin{abstract}
We present a Monte Carlo study of the high-temperature 
phase of the two-dimensional driven lattice gas 
at infinite driving field. We define a finite-volume 
correlation length, verify that this definition has a good 
infinite-volume limit independent of the lattice geometry, and 
study its finite-size-scaling behavior.  
The results for the correlation length are
in good agreement with the predictions based on the field theory
proposed by Janssen, Schmittmann, Leung, and Cardy.
The theoretical predictions for the susceptibility and 
the magnetization are also well verified.  
We show that the transverse Binder parameter vanishes at the 
critical point in all dimensions $d\ge 2$ and discuss how such 
result should be expected in the theory of Janssen {\em et al.} 
in spite of the existence of a dangerously irrelevant operator.
Our results confirm the Gaussian nature of the transverse excitations.
\end{abstract}

\clearpage

%
%

\newcommand{\reff}[1]{(\ref{#1})}
\newcommand{\kls}{{\sc KLS}}
\newcommand{\su}[2]{\frac{#1}{#2}}
\newcommand{\me}{{\it master equation}}
\newcommand{\db}{{\it detailed balance}}
\newcommand{\ve}[1]{{\bf #1}}

\newcommand{\dlg}{{\sc DLG}}
\newcommand{\fss}{{\sc FSS}}

\newcommand{\pe}[1]{#1_{\bot}}
\newcommand{\pa}[1]{#1_{\|}}
\newcommand{\spint}[1]{\ensuremath{\int {\rm d}^{#1}\ve{x} \, {\rm d}t}}
\newcommand{\imint}[2]{\ensuremath{\int \! \frac{ {\rm
        d}^{#1}\ve{#2}}{(2\pi)^{#1}} \, \frac{{\rm d}\omega}{2\pi}}} 
\newcommand{\nbpa}{\ensuremath{\Delta_{\|}}}
\newcommand{\nbpe}{\ensuremath{\Delta_{\bot}}}
\newcommand{\grpe}{\ensuremath{\vett{\nabla}_{\bot}}}
\newcommand{\grpa}{\ensuremath{\nabla_{\|}}}
\newcommand{\qpeq}{\ensuremath \pe{\ve{q}}^{2}}
\newcommand{\ts}{\ensuremath \tilde{s}}
\newcommand{\G}[2]{\ensuremath \Gamma_{#1\,#2}} 
\newcommand{\tn}{\ensuremath \tilde{n}}

\section{Introduction}

At present, the statistical mechanics of systems in thermal equilibrium is 
quite well established. On the other hand, little is known in general
for nonequilibrium systems, although some interesting results have been
recently obtained \cite{ref-generali}. 
It seems therefore worthwhile to 
study simple models which are out of thermal equilibrium. 
One of them was introduced
at the beginning of the eighties by 
Katz, Lebowitz, and Spohn~\cite{Katz}, who studied the 
stationary state of a lattice gas under the action of an external
drive. The model, hereafter called driven lattice gas (\dlg), is
a kinetic Ising model on a periodic domain with Kawasaki dynamics and
biased jump rates. Although not in thermal equilibrium, the \dlg\
has a time-independent stationary state and shows a finite-temperature 
phase transition, which is however different in nature from its
equilibrium counterpart.\footnote{For an extensive presentation of the \dlg\
and of many generalizations, see Refs. \cite{Schmittmann95,MarroDickman}.}

Despite its simplicity, the \dlg\  has not yet been solved exactly\footnote{
The \dlg\ is soluble for infinite drive in the limit in which jumps 
in the direction of the field are infinitely more frequent than 
jumps in the orthogonal directions \cite{vB-S-84}.}
and at present there is still much debate on the nature of the phase
transition \cite{Janssen86a,Leung86,Garrido98,Garrido99,Garrido00,Viability99,%
leung00}.  In Refs.~\cite{Janssen86a,Leung86}
Janssen, Schmittmann, Leung, and Cardy (JSLC)  
developed a continuum theory which should  
capture the basic features of the transition and which provides 
exact predictions for the critical exponents.
Several computer simulations in two 
and three dimensions~\cite{Zhang88,Leung92,Wang96,Leung99}
provided good support to these field-theoretical predictions, once it was
understood that the highly anisotropic character of the transition
required some kind of anisotropic finite-size scaling (FSS) 
\cite{Binder89,Leung92}. 
Still some discrepancies remained, prompting Garrido, de los Santos,
and Mu\~{n}oz \cite{Garrido98,Garrido99,Garrido00}
to reanalyze the derivation of the field theory. 
On the basis of this analysis, they suggested that the DLG at 
infinite driving field should not behave as predicted by JSLC 
but should rather belong to the universality class of the randomly driven
lattice gas (RDLG) \cite{SZ-91,Schmittmann-93}. This approach gives different 
predictions for the critical exponents that have been apparently
verified numerically \cite{AGMM,AS-02}, see also Ref.~\cite{Munoz-02}. 

In view of these contradictory results, a new numerical investigation 
is necessary, in order to decide which theory describes the critical 
behavior of the DLG. For this purpose it
is useful to consider quantities that are exactly predicted at least in one 
of the two theories. Here we shall focus on transverse 
fluctuations since in the JSLC theory transverse correlation functions
are predicted to be Gaussian.
Therefore, beside critical exponents, one can also exactly compute 
the FSS functions of several observables
and make an unambiguous test of the JSLC theory.         

A basic ingredient of our FSS analysis is the finite-volume correlation length.
In spite of the extensive numerical work, 
no direct studies of the correlation 
length have been done so far, essentially because it is not easy to define it.
Indeed, in the high-temperature phase the model
shows long-range correlations due to the violation of detailed
balance~\cite{Garrido90,Grinstein91}. 
Therefore, no correlation length can be defined from
the large-distance behavior of the two-point correlation function. A 
different definition is therefore necessary: In Refs. \cite{Valles8687}
a parallel correlation length is defined. However, this definition suffers
from many ambiguities (see the discussion in Ref. \cite{Schmittmann95}) 
and gives results for the exponent $\pa{\nu}$ which are not in agreement 
with the JSLC theory \cite{Zhang88}. Even more difficult appears the definition
of a transverse correlation length because of the presence of negative 
correlations at large distances \cite{Zhang88,Schmittmann95}.

In this paper we define a finite-volume transverse correlation length
generalizing the definition of the second-moment correlation length that is 
used in equilibrium systems. Because of the conserved dynamics, such a 
generalization requires some care. Here, we use the results of 
Ref.~\cite{correlation_length}.
The main advantage of using a correlation length is the possibility 
of performing FSS checks without free parameters. 
We find that the susceptibility and the correlation length show a 
FSS behavior that is in full agreement with the idea that transverse 
fluctuations are Gaussian. This immediately implies the JSLC
predictions $\gamma_\bot = 1$ and $\nu_\bot=1/2$.
We also study the FSS behavior of the magnetization,
finding $\beta_\bot/\nu_\bot = 1.023(43)$, 
in agreement with mean-field behavior.
Finally, we consider the transverse Binder parameter and find that 
it goes to zero at $\beta_c$ as the volume increases. 
This supports again the Gaussian nature of the transverse mode and, 
as we discuss in detail, it 
is not in contradiction with the presence 
of a dangerously irrelevant operator. Indeed, since no zero mode is present
in the theory, there should be no anomalous scaling. 

The paper is organized as follows.
In Sec.~\ref{sec-2} we describe the model and define the
observables measured in the Monte Carlo simulation.
Section~\ref{sec:fss} reviews the FSS theory and introduces the 
basic formulae that are used in the analysis
of  the numerical data.  Next in Sec.~\ref{sec-4}, we consider the 
field theory of Refs.~\cite{Janssen86a,Leung86} in a finite 
geometry and compute the FSS functions for several observables. 
In Sec.~\ref{sec:simulations} we describe the
simulations and present the results which are then discussed 
and compared with recent findings \cite{AGMM,AS-02}
in Sec.~\ref{sec:conclusions}. 
We confirm the field-theoretical predictions of JSLC 
both for infinite-volume quantities and for the finite-size behavior.
In App.~\ref{AppA} we consider the $O(N)$ model for $N\to \infty$ above the 
upper critical dimension and discuss the critical behavior of different 
definitions of the Binder parameter. In particular, we show that, if 
it is defined in terms of correlation functions at nonzero 
momenta, then it vanishes for all $d\ge 4$: this is the same behavior
as that expected in the DLG for $d \ge 2$ on the basis of the JSLC theory 
and verified numerically  for $d=2$. In App.~\ref{AppB} we sketch a 
one-loop calculation of the Binder parameter in the JSLC theory.
A short account of the results presented in this work has been given in 
Ref.~\cite{CGGP-02}.

\section{Definitions}
\label{sec-2}

\subsection{The model}
\label{sec-2.1}

We consider a finite square lattice $\Lambda$\ 
and $N$\ particles, each of them occupying a different lattice site. A
configuration of the system is specified by the set of
occupation numbers of each site 
$ n = \{n_{i} \in \{0,1\}\}_{i\in\Lambda}.  $
The standard lattice gas is characterized by 
a nearest-neighbor attractive (``ferromagnetic'' in spin language)
Hamiltonian 
\begin{equation}
        H_{\Lambda}[n]=-4\sum_{\langle i,j \rangle\in\Lambda} n_{i}\,n_{j},
\end{equation} 
where the sum runs over all lattice nearest neighbors.
We consider a  discrete-time Kawasaki dynamics~\cite{Kawasaki},
which preserves the total number of particles $N$ or, 
equivalently, the density 
\begin{equation}
        \rho_{\Lambda} \equiv  \su{1}{|\Lambda|}\sum_{i\in\Lambda} n_{i},
\end{equation}
where $|\Lambda|$\ is the total number of sites in $\Lambda$.
At each step, we randomly choose a lattice link $\langle i,j \rangle$. 
If $n_i = n_j$, nothing happens. Otherwise, we propose 
a particle jump with probability $w(\Delta H/T)$, where 
\begin{equation}
        \Delta H = H_\Lambda[n'] - H_\Lambda[n]
\end{equation}
is the difference in energy between the new ($n'$) and the old ($n$) 
configuration. If the probability $w(x)$ satisfies
\begin{equation}
        w(-x)=e^{x}\,w(x),
\end{equation} 
then the dynamics is reversible, i.e. satisfies detailed balance.
Under these conditions there is a unique equilibrium measure
given by 
\begin{equation}
        P_{\Lambda,eq}[n]=\su{e^{-\beta
        H_\Lambda[n]}}{\sum_{\{n'\}}\,e^{-\beta H_\Lambda[n']}} ,
\end{equation} 
where $\beta \equiv 1/T$.
In the thermodynamic limit
the lattice gas exhibits a second-order phase transition 
for $\rho_\Lambda = 1/2$ and \mbox{$\beta_{c}=\su{1}{2}\ln{ (\sqrt{2}+1)}$}, 
which belongs to the standard Ising universality class.

The DLG is a generalization of the lattice gas in which one 
introduces a uniform (in space and time) force field
pointing along one of the axes of the lattice, i.e. $\ve{E}=E\ve{\hat{x}}$:
It favors (respectively suppresses) the jumps of the particles in the 
positive (resp. negative) $\ve{\hat{x}}$-direction.
If $\Lambda$\ is
bounded by {\it rigid walls}, then $\ve{E}$\ is a conservative field and 
it can be accounted for by adding a potential term
to $H_\Lambda[n]$. Therefore, the system remains in thermal equilibrium.
The net effect of $\ve{E}$ is simply to induce a concentration gradient in the
equilibrium state. 

Here, we consider instead periodic boundary conditions.\footnote{
In principle, it is enough to consider periodic boundary conditions in the
field direction. The boundary conditions in the transverse directions
are largely irrelevant for the problems discussed here. }
In this case, the field $\ve{E}$ does not have a global potential and  
the system reaches a stationary state which, however, is not a state 
in thermal equilibrium.

In the \dlg\  transition
probabilities take into account the work done by the field during the
particle jump from one site to one of its nearest neighbors. In this case
one proposes a particle jump with probability 
       $w(\beta\,\Delta H+\beta E \ell )$     ,
with $\ell = (-1, 0, 1)$\ for jumps (along, transverse, opposite) to 
$\ve{\hat{x}}$.

For $E \neq 0$\ and $\rho_{\Lambda}=1/2$, the system 
undergoes a continuous phase
transition \cite{Schmittmann95,MarroDickman} at an
inverse temperature $\beta_{c}(E)$\ which saturates, for
$E\rightarrow\infty$, at $\beta_{c}(\infty) \approx 0.71
\beta_{c}(0)$. For $\beta<\beta_{c}(E)$\ particles are  homogeneously
distributed in space, while for $\beta>\beta_{c}(E)$\,  phase separation
occurs:
Two regions are formed, one almost full and the other one almost empty,
with interfaces parallel to $\ve{E}$.

\subsection{Observables}
\label{sec:observables}

We consider a finite square lattice of size $\pa{L}\times
\pe{L}$ with periodic boundary conditions. We define 
a ``spin" variable $s_{\ve{j}} \equiv  2 n_{\ve{j}} - 1$ 
and its Fourier transform 
\begin{equation}
        \phi(\ve{k}) \equiv \sum_{\ve{j}\in\Lambda} 
         e^{i \ve{k}\cdot\ve{j}} s_{\ve{j}},
\end{equation}
where the allowed momenta are
\begin{equation}
        \ve{k}_{n,m} \equiv
      \left (\su{2\pi n}{\pa{L}},\su{2\pi m}{\pe{L}} \right ),
\end{equation}
with $(n,m)\in\mathbb{Z}_\protect{\pa{L}}\times\mathbb{Z}_\protect{\pe{L}} $. 

We consider the model at half filling, i.e. for $\rho_\Lambda = 1/2$. Then
\begin{equation}
        \sum_{\ve{j}\in\Lambda} s_{\ve{j}} = 0, \qquad\qquad 
        {\rm i.e.} \qquad\qquad
        \phi(\ve{k}_{0,0})=0. \label{zero}
\end{equation}
In the ordered phase $|\phi(\ve{k})|$\ takes its maximum for \mbox{$\ve{k} =
\ve{k}_{0,1}$}, and the expectation value on the steady state of its module
\begin{equation}
        m(\beta;\pa{L},\pe{L}) \equiv {1\over |\Lambda|}
            \langle|\phi(\ve{k}_{0,1})| \rangle
\label{magnetization}
\end{equation}
is a good order parameter.

In momentum space the static structure factor, the Fourier transform of the 
two-point correlation function,
\begin{equation}
        \tilde{G}(\ve{k}; \pa{L},\pe{L})\equiv 
        {1\over |\Lambda|} \langle|\phi(\ve{k})|^{2}\rangle
\label{statstrfa}
\end{equation}
vanishes at $\ve{k}_{0,0}$\ because of Eq.~(\ref{zero}) and attains 
its maximum at
$\ve{k}_{0,1}$, so that it is natural to define the susceptibility
as\footnote{We note that the susceptibility  defined by using the
  linear response theory does not coincide in nonequilibrium systems
  with that defined in terms of the Fourier transform of the two-point
  correlation function. }
\begin{equation}
        \chi_\bot(\beta;\pa{L},\pe{L})\equiv 
    \tilde{G}(\ve{k}_{0,1}; \pa{L},\pe{L}).
\label{defChi}
\end{equation}
We also define the four-point connected correlation function
\begin{equation}
\tilde{G}^{(4)}(\ve{k}_1, \ve{k}_2, \ve{k}_3, \ve{k}_4; L_\|,L_\bot) = 
   {1\over |\Lambda|} 
  \< \phi(\ve{k}_1) \phi(\ve{k}_2) \phi(\ve{k}_3) \phi(\ve{k}_4)\>_{\rm conn}
  \; ,
\end{equation}
and the related transverse Binder cumulant $g(\beta;\pa{L}, \pe{L})$
defined as
\begin{equation}
\label{defbinder}
g(\beta;\pa{L}, \pe{L})  \equiv  2- 
\frac{\langle|\phi(\ve{k}_{0,1})|^{4}\rangle}
{\langle|\phi(\ve{k}_{0,1})|^{2}\rangle^2} = 
  - {\tilde{G}^{(4)}(\ve{k}_{0,1}, \ve{k}_{0,1}, 
         -\ve{k}_{0,1}, -\ve{k}_{0,1}; L_\|,L_\bot) \over 
     |\Lambda| [\tilde{G}(\ve{k}_{0,1};L_\|,L_\bot)]^2}.
\end{equation}
Next, we would like to define a correlation length. In infinite-volume 
equilibrium systems there are essentially two different ways 
of doing it. One can define the correlation length in terms of the 
large-distance behavior of the two-point correlation function or by using its
small-momentum behavior (second-moment correlation
length). In the DLG the first method does not work. 
Indeed, in the high-temperature phase the two-point correlation function 
always decays {\it algebraically} with the distance.
Moreover, it is not positive definite because of
negative correlations in the transverse directions~\cite{Schmittmann95}. 
This peculiar behavior is due to the fact
that in the infinite-volume limit (at fixed temperature)
the static structure factor 
$\tilde{G}(\ve{k};\infty,\infty)$ has a finite discontinuity at $\ve{k}=0$. 

In this paper we propose a new definition that generalizes the 
second-moment correlation length used in equilibrium spin
systems. The basic observation is that in the DLG the infinite-volume 
{\em wall-wall} correlation function decays exponentially, i.e. 
\begin{equation}
\sum_{x_\|} G((x_\|,x_\bot);\infty,\infty) \equiv 
   \int d^{d-1}q_\bot\, \tilde{G}((0,q_\bot);\infty,\infty) 
   e^{iq_\bot\cdot x_\bot} \sim e^{-\kappa |x_\bot|},
\label{aaa}
\end{equation}
as in equilibrium systems. This holds at tree level 
both in the JSLC and in RDLG field theories and to all orders of 
perturbation theory in the JSLC theory, see Sec.~\ref{sec-4}.
Therefore, we can define a correlation length as in equilibrium systems,
paying due attention to the conserved dynamics.
Here, we follow
Ref. \cite{correlation_length}, where we discussed the possible 
definitions of correlation length in the absence of the zero mode, 
as it is the case here. 

We consider the structure factor in finite volume
at zero longitudinal momenta
\begin{equation}
        \pe{\tilde{G}}(q; \pa{L}, \pe{L}) \equiv\, 
           {\tilde{G}}((0,q); 
        \pa{L}, \pe{L}) ,
\label{struct-bot}
\end{equation}
(note that the conservation law 
implies $\pe{\tilde{G}}(0; \pa{L}, \pe{L})=0$)
and define a finite-volume (transverse) correlation length\footnote{
In Ref.~\cite{correlation_length} we showed that any good 
finite-volume correlation length must satisfy two properties:
(i) it must be finite for all $T\not=0$ and $L_\bot,L_\| < \infty$;
(ii) it must diverge as $T\to 0$ even in finite volume. 
The definition (\ref{defxi}) satisfies these two properties. 
Note that in infinite volume one can also define a correlation length
from the large-distance behavior of the correlation function, i.e.
one can define $\xi_{\infty,\bot} = 1/\kappa$, where $\kappa$ is defined 
in Eq.~(\ref{aaa}). Such a definition is not convenient here, since it does not 
admit a finite-volume generalization.
}
\begin{equation}
\label{defxi}
        {\xi}_{ij}(\pa{L}, \pe{L}) \equiv
        \sqrt{\su{1}{\hat{q}_{j}^{2} - \hat{q}_{i}^{2}}
        \left( 
\su{\pe{\tilde{G}}(q_{i}; \pa{L}, \pe{L})}{
        \pe{\tilde{G}}(q_{j}; \pa{L},
        \pe{L})} -1\right) },
\end{equation}
where $\hat{q}_{n}=2\sin{(\pi n/\pe{L})}$\ is the lattice momentum.
Since $\pe{\tilde{G}}(0; \pa{L}, \pe{L})=0$,
$q_i$ and $q_j$ must not vanish. Moreover, 
as discussed in Ref. \cite{correlation_length}, the definition
should be valid for all $\beta$ in finite volume. Since the system orders in an 
even number of stripes, for $i$ even $\pe{\tilde{G}}(q_i; \pa{L}, \pe{L})$ 
is zero as $\beta\to \infty$. Therefore, if our definition should capture 
the nature of the phase transition, we must require $i$ and $j$ to be odd.
Although any choice of $i,j$ is conceptually good, 
finite-size corrections increase with $i,j$, a phenomenon which should 
be expected since the critical modes correspond to 
$q\to 0$. Thus, we choose $(i,j)$ = $(1,3)$, defining 
$\xi_\bot \equiv \xi_{13}$.

The same method can be used to define a longitudinal correlation 
length, although in this case we do not have an all-order proof 
(not even in the JSLC theory) that wall-wall longitudinal correlations
decay exponentially. 
It is enough to consider the longitudinal structure factor 
at zero transverse momentum 
\begin{equation}
\tilde{G}_\| (q;L_\|,L_\bot) = \tilde{G}((q,0);L_\|,L_\bot),
\end{equation}
and use again Eq.~(\ref{defxi}). In this case, there is no reason to 
avoid the use of even $i$ or $j$ and thus we define 
$\xi_\| \equiv \xi_{12}$.

Near a phase-transition point the quantities we have defined above 
show power-law divergences. As usual, see, e.g., Ref.~\cite{Schmittmann95},
we define transverse exponents by assuming
\begin{eqnarray}
\xi_\bot &\sim& t^{-\nu_\bot} \nonumber \\
\chi_\bot &\sim& t^{-\gamma_\bot} , 
\end{eqnarray}
for $t \equiv (\beta_c - \beta)/\beta_c\to 0^+$. 
The magnetization vanishes in the low-temperature phase as
\begin{equation}
m \sim (-t)^{\beta_\bot}.
\end{equation}

\section{Finite-size scaling}
\label{sec:fss}

In the neighborhood of a critical point 
the behavior of long-range observables is controlled
by few quantities, corresponding in the renormalization-group language
to coordinates parametrizing the relevant directions in the 
infinite-dimensional coupling space.
When the system is finite, its size plays the role
of another relevant operator. This means that an observable
$\mathcal{O}$, which diverges in the thermodynamic limit as
\be
\mathcal{O}_\infty(\beta) \sim t^{-\gamma_\mathcal{O}} \qquad
\hbox{for } t \equiv 1-{\beta \over \beta_c} \to 0^+ ,
\end{equation}
behaves in a finite system of size $L_\|\times L_\bot^{d-1}$ as
\cite{Fisher-71,Barber}
\be
\mathcal{O}(\beta;L_\|,L_\bot) \approx t^{-\gamma_\mathcal{O}}
f_{1,\mathcal{O}}(t^{-\nu} /L_\bot;S) \approx
L_\bot^{\gamma_\mathcal{O}/\nu} f_{2,\mathcal{O}}(t^{-\nu} /L_\bot;S) \approx 
L_\bot^{\gamma_\mathcal{O}/\nu} f_{3,\mathcal{O}}(\xi_\infty(\beta)/L_\bot;S),
\label{FSS-basic}
\end{equation}
where $\xi_\infty(\beta)$ is the infinite-volume correlation length 
and $S \equiv L_\|/L_\bot$ is the aspect ratio that is kept fixed in the 
FSS limit. From Eq.~\reff{FSS-basic} we can derive a general relation for
the ratio of $\mathcal{O}(\beta;L_\|,L_\bot)$ at two different sizes 
$(L_\|, L_\bot)$ and $(\alpha L_\|, \alpha L_\bot)$.
In the FSS limit we obtain
\begin{equation}
\label{eq:fssform1}
{\mathcal{O}(\beta;\alpha L_\|, \alpha L_\bot) \over 
 \mathcal{O}(\beta;L_\|, L_\bot)} = F_\mathcal{O} \left
  (\alpha,  \frac{\xi(\beta;L_\|, L_\bot)}{L_\bot}, S\right)  ,
\end{equation}
where we have replaced $\xi_\infty(\beta)/L_\bot$ with 
$\xi(\beta;L_\|, L_\bot)/L_\bot$ by inverting
$\xi(\beta;L_\|, L_\bot) \approx L_\bot f_{3,\xi}(\xi_\infty(\beta)/L_\bot;S)$.
The function $F_\mathcal{O}(\alpha,z,S)$ is universal
and is directly accessible numerically, 
e.g., by Monte Carlo simulations---no need to fix any parameter---since all 
quantities appearing in Eq.~(\ref{eq:fssform1}) are directly measurable.
Moreover, as we shall show 
below, all critical exponents can be determined from the 
FSS functions $F_\mathcal{O}(\alpha,z,S)$ independently of the critical 
temperature.

If we define $z \equiv \xi(\beta;L_\|, L_\bot)/L_\bot$, 
then $z$ varies between 0 and $z^*(S)$, where $z^*(S)$ is defined by 
\be
 z^*(S) = f_{3,\xi} (\infty,S),
\label{def-zstar}
\end{equation}
or implicitly from 
\begin{equation}
 \alpha = F_\xi(\alpha,z^*(S),S).
\end{equation}
The value $z^*(S)$ is directly related to the behavior of the finite-size
correlation length at the critical point, since 
$\xi(\beta_c;L_\|, L_\bot) \approx z^*(S) L_\bot$. 
For finite-temperature phase transitions $z^*(S)$ is finite. 
By considering the behavior of the FSS functions at 
$z^*(S)$ we can determine the exponents $\gamma_{\cal O}/\nu$. Indeed, 
at the critical point we have
\begin{equation}
\mathcal{O}(\beta_c;L_\|, L_\bot) \sim L_\bot^{\gamma_{\mathcal{O}}/\nu},
\label{eq:critical_scaling}
\end{equation}
so that 
\be
F_{\mathcal{O}}(\alpha,z^*(S),S) =  
   \frac{\mathcal{O}(\beta_c;\alpha L_\|, \alpha L_\bot)}{
         \mathcal{O}(\beta_c;L_\|, L_\bot)} = 
    \alpha^{\gamma_{\mathcal{O}}/\nu},
\end{equation}
and therefore 
\begin{equation}
\frac{\gamma_{\mathcal{O}}}{\nu} = 
\frac{\log F_{\mathcal{O}}(\alpha,z^*(S),S)}{\log \alpha}.
\label{eq:dummy_dimensions}
\end{equation}                                                                  
By studying the behavior of $F_\xi(\alpha,z,S)$ in a neighborhood of 
$z^*(S)$ it is also possible to derive the exponent $\nu$. Using the 
fact that 
\begin{equation}
{\xi(\beta;L_\|, L_\bot)\over L_\bot} \approx 
   z^*(S) + a(S) (\beta - \beta_c) L^{1/ \nu}_\bot,
\end{equation}
near the critical point, we obtain 
\begin{equation} 
  \left. z {dF_\xi(\alpha,z,S)\over dz}\right |_{z = z^*(S)} = 
  \alpha (\alpha^{1/\nu} - 1).
\label{nu-da-F}
\end{equation} 
The above-presented results are valid for an isotropic system. 
On the other hand, the numerical simulations and the field-theoretical
studies predict that the phase transition in the \dlg\
is strongly anisotropic.  For example, the scaling form of the
critical static two-point function should be 
\be
\tilde{G}(k_\|,k_\bot) \approx \mu^{-2+\eta_\bot}
\tilde{G}(\mu^{1+\Delta} k_\|, \mu k_\bot),
\end{equation}
where $\eta_\bot$ is the anomalous dimension of the density field 
(see Ref.~\cite{Schmittmann95} for definitions) and 
$\Delta$ is the so-called \emph{anisotropy exponent}.

It is then natural to assume the existence of two correlation lengths
$\xi_\bot, \xi_\|$ which diverge with different  exponents $\nu_\bot$ and
$\nu_\|$ related by \cite{Schmittmann95}
\be
\nu_\| = (1+\Delta) \nu_\bot.
\end{equation}

These considerations call for an extension of the FSS arguments.
A phenomenological approach to \fss\ for the \dlg\ has been
developed~\cite{Valles8687}, keeping into account the strong
anisotropy observed in the transition (for $d=2$  see Refs.~\cite{Wang96,%
Leung92,Binder89}, for $d=3$ see Ref.~\cite{Leung99}).  
Following this approach, we 
assume that all observables have a finite FSS limit
for $\pa{L}$, $\pe{L}\to\infty$ keeping constant:
\begin{itemize}
\item the \emph{anisotropic aspect ratio} 
$
S_\Delta \equiv \pa{L}^{1/(1+\Delta)}/\pe{L};
$ 
\item the \emph{FSS parameter}
$\xi_{\bot,\infty}(\beta)/\pe{L}$ (or equivalently its
longitudinal counterpart).
\end{itemize}
Then, Eq. \reff{FSS-basic} still holds by using the correct parameters,
i.e. by replacing $S$ with $S_\Delta$, $\nu$ with $\nu_\bot$, and 
$\xi_\infty$ with $\xi_{\bot,\infty}$.
Analogously Eq.~(\ref{eq:fssform1}) is recast in the form
\begin{equation}
\label{eq:fssform2}
{\mathcal{O}(\beta;\alpha^{1+\Delta} L_\|, \alpha L_\bot) \over 
      \mathcal{O}(\beta; L_\|,L_\bot)}
      = F_\mathcal{O} \left (\alpha,
      \frac{\xi_\bot(\beta;L_\|,L_\bot)}{\pe{L}},S_\Delta\right).   
\end{equation}
Equation (\ref{eq:fssform2}) is the basis for our analysis of the phase
transition in the \dlg\ .  For the transverse finite-volume correlation 
length $\xi_\bot$ we use $\xi_{13}$ defined in Sec. \ref{sec:observables}.

In anisotropic systems the FSS limit must be taken at fixed $S_\Delta$, 
which in turn requires the knowledge of the exact value of $\Delta$. It is 
thus important to understand how we can single out the correct value 
of $\Delta$ from the simulations. First, it should be noticed
that observing FSS does not imply that we are using the correct value of 
$\Delta$ \cite{CGGP-shape}. 
Indeed, note first that Eqs.~(\ref{FSS-basic}) and 
(\ref{eq:fssform2}) still hold for $L_\|=\infty$---this corresponds to 
$S_\Delta=\infty$---i.e. for a geometry 
$\infty\times L_\bot^{d-1}$ (in two dimensions this is a strip). 
Then, imagine that we use an incorrect value $\delta$, keeping $S_\delta$ 
fixed in the FSS limit. If $\delta > \Delta$, $L_\|$ increases 
much faster than it should and $S_\Delta\to\infty$. We thus expect to obtain 
the FSS behavior corresponding to a geometry $\infty\times L_\bot^{d-1}$.
Thus, we should be able to observe scaling whenever we use $L_\bot$ as 
reference length and consider transverse quantities. 
On the other hand, longitudinal quantities are expected not to scale properly.
For instance, $\xi_\|/L_\|$ should not have a good FSS behavior when plotted 
vs $\xi_\bot/L_\bot$. 
Indeed, at fixed $S_\Delta$ we have 
that $\xi_\| \sim L_\| \sim L_\bot^{1 + \Delta}$. If instead $S_\delta$ is 
fixed and $\delta > \Delta$, 
$L_\|$ increases too fast, so that we expect $\xi_\|$ to be controlled
by the transverse size only. Thus, $\xi_\|$ should increase slower than
$L_\|$, and thus, at fixed $\xi_\bot/L_\bot$, one should observe 
$\xi_\|/L_\|\to 0$. If $\delta < \Delta$ the same argument holds by simply
interchanging longitudinal and transverse quantities.
This observation provides therefore a method to determine the 
correct value of $\Delta$. It is the value for which both correlation lengths
scale correctly, i.e. $\xi_\| \sim L_\|$ and $\xi_\bot \sim L_\bot$.

\section{Field-theory description of the DLG} \label{sec-4}

As we explained in Sec. \ref{sec-2.1}, the \dlg\ is a lattice gas
model. However, in a neighborhood of the critical
point (critical region) we can limit ourselves to consider 
slowly-varying (in space and time) observables.
At criticality the lattice spacing is
negligible compared to the length and time scales at which long-range
order is established so that it is possible to formulate a
description of the system in terms of {\it mesoscopic} variables.
In principle, the dynamics of such variables
can be obtained by coarse graining the microscopic system.
However, given the difficulty of performing a rigorous coarse-graining
procedure, one postulates a continuum field theory 
that possesses all the symmetries of the microscopic lattice model. 
By universality the continuum theory should have the same critical behavior 
of the microscopic one. 

Unfortunately, there is at present no consensus on the field theory that 
describes the critical behavior of the DLG.\footnote{
The effects of an external drive on the standard Model-B dynamics
were also studied in Ref.~\cite{GawKupi86}.}
The theory originally 
proposed by JSLC~\cite{Leung86,Janssen86a} has been recently disputed in 
Refs.~\cite{Garrido98,Garrido99,Garrido00} (see also
Ref.~\cite{Viability99}), where it is proposed that the DLG 
at infinite driving field is in the same universality class of 
the randomly driven lattice gas (RDLG) \cite{SZ-91,Schmittmann-93}. 
Both theories agree on the anisotropic nature of the 
phase transition, but disagree on its origin and make different predictions
for the universal quantities.

In the JSLC theory the critical exponents are exactly predicted, and, for 
$2 \le d \le 5$, they are given by
\begin{eqnarray}
\eta_\bot &=& 0, \\
\nu_\bot  &=& {1\over2}, \label{nu-theory} \\
\gamma_\bot  &=& {1}, \label{gamma-theory} \\
\beta_\bot  &=& {1\over2}, \label{beta-theory} \\
\Delta &=& {1\over3} (8 - d). \label{delta-theory}
\end{eqnarray}
In the RDLG model critical exponents are only known up to two loops 
perturbatively in $\epsilon \equiv 3 - d$. Therefore,
it is difficult to estimate how much they differ from the JSLC ones. 
We note that 
$\Delta$ should be quite different in the two theories in two dimensions.
In the RDLG we should have $\Delta \approx 1$,
since \cite{Schmittmann-93} $\Delta = 1 - \eta/2$
and we expect $\eta$ to be small. On the other hand, 
Eq.~(\ref{delta-theory}) gives $\Delta = 2$.

For the JSLC theory, not only do we have exact predictions for the exponents,
but we can also compute exactly the transverse structure factor 
(\ref{struct-bot}).
Keeping into account causality 
\cite{Janssen86a,Janssen76,Bausch76,Janssen79}
and the form of the interaction vertex
one can see that for $\pa{k} = 0$  there are no loop contributions to the 
two-point function $\tilde{G}(k)$ (and also to the two-point response function).
Thus, for all $2\le d\le 5$, $\tilde{G}_\bot (k)$ is simply given, 
in the field-theoretical approach of JSLC,
by the tree-level expression
\be
\tilde{G}_\bot (k) = {1\over k^2 + \tau},
\label{sec4:tildeG}
\end{equation}
where $\tau$ is a squared ``bare mass" that vanishes at criticality.
Two observations are in order. First, $\tau =  b t + O(t^2)$ 
for $t\equiv (\beta_c - \beta)/\beta_c\to 0$ with $b$ positive constant. 
Second, the function that appears in 
Eq.~\reff{sec4:tildeG} refers to the coarse-grained fields, which, in 
the critical limit, differ by a finite renormalization from the lattice ones. 
Thus, for the lattice function we are interested in, in the scaling 
limit $t\to 0$, $k\to 0$, with $k^2/t$ fixed,
we have
\be
\tilde{G}_{\bot,\rm latt} (k) = {Z\over k^2 + b t},
\label{sec4:tildeG-lattice}
\end{equation}
where $Z$ and $b$ are positive constants.

On the same footing, we can conclude that all correlation functions 
with vanishing longitudinal momenta behave as in a free
theory. In particular, the Binder cumulant defined in  Eq.~(\ref{defbinder})
vanishes. 
It is also important to notice that Eq.~\reff{sec4:tildeG-lattice} implies the 
exponential decay of 
\be
    {G}_{\bot,\rm latt} (\pe{x}) = \int d^{d-1}k \, e^{ik\pe{x}} \
      \tilde{G}_{\bot,\rm latt} (k), 
\end{equation}
which fully justifies our definition of transverse correlation 
length. 

Field-theoretical methods can also be used to determine the 
FSS behavior of the model.
Predictions are easily obtained by following the 
method applied in equilibrium spin systems 
(see, e.g., Ref. \cite[Chap.~36]{ZinnBook}
 and references therein). The idea  is quite simple. 
Consider the system in a finite box with periodic boundary conditions. 
The finite geometry has the only effect of quantizing the momenta. 
Thus, the perturbative finite-volume correlation functions are obtained 
by replacing momentum integrals by lattice sums. Ultraviolet divergences
are not affected by the presence of the box~\cite{Brezin82} and thus
one can use the infinite-volume renormalization constants.
Once the renormalization is carried out, one obtains the 
geometry-dependent finite-size correlation functions.

The considerations we have presented above for the infinite-volume case
apply also in finite volume and thus Eq.~(\ref{sec4:tildeG-lattice}) 
holds in this case.
Using Eq.~(\ref{sec4:tildeG-lattice}), in the FSS limit we find
\begin{eqnarray}
\label{fssxi}
        \frac{{\xi}_\bot(\beta;\pa{L},\pe{L})}{L_\bot} &=& 
        \left[ (2\pi)^2 +  b t \pe{L}^2 \right]^{-1/2}, 
\end{eqnarray}
valid for $t\to 0$, $\pa{L},\pe{L}\to \infty$ with $t \pe{L}^2$ fixed.
Although we have not explicitly mentioned $S_\Delta$ and 
such a quantity does not appear in 
Eq.~(\ref{fssxi}), this expression is 
expected to be valid only if the FSS limit is taken by keeping 
$S_\Delta$ constant.
Taking the limit $\pe{L}\to\infty$ we obtain 
$\xi_{\bot,\infty}(\beta)^2 \approx  1/(b t)$,
so that we can write Eq.~(\ref{fssxi}) also in the form
\begin{equation}
\label{fssxi2}
{1\over [{\xi}_\bot(\beta;\pa{L},\pe{L} )]^2} =
      {1\over \xi_\infty(\beta)^2} + {4\pi^2\over L_\bot^2} \; .
\end{equation}
Using Eqs.~(\ref{fssxi2}) and (\ref{sec4:tildeG-lattice})
 we can also compute the scaling functions 
$F_\xi(\alpha,z,S_\Delta)$ and $F_\chi(\alpha,z,S_\Delta)$ 
defined in Eq. \reff{eq:fssform2}. We obtain
\begin{eqnarray}
        F_{\xi}(\alpha,z,S_\Delta) &=& 
        \left[ 1- \left(1 -\alpha^{-2}\right)(2\pi)^2 z^2
         \right]^{-1/2},
\label{eq:theoretical_pred} \\
F_{\chi} (\alpha,z,S_\Delta) &=& F_\xi^2(\alpha,z,S_\Delta) = 
   \left[1- \left(1 -\alpha^{-2}\right)(2\pi)^2 z^2\right]^{-1}.
\label{eq:theoretical_pred_chi} 
\end{eqnarray}
Note that these functions do not depend on $S_\Delta$ and that 
\begin{equation}
z^*(S_\Delta) = {1\over 2\pi} \; .
\end{equation}
A peculiarity of the JSLC theory 
is the presence of an operator with renormalization-group dimension 
$2\sigma = 2(d-2)/3$ that is dangerously irrelevant
for $2 < d < 5$ and becomes marginal at $d=2$. 
Keeping into account the coupling $u$ associated with the dangerously 
irrelevant operator, in the JSLC theory we have \cite{Janssen86a}
\begin{eqnarray}
&& \Gamma_{\tilde n,n}(\{(q_{\|}, q_\bot )\},\omega,\tau,u;L_\|, L_{\bot}) = 
\label{scaling-Gamma-nnt} \\
&& \qquad = 
\ell^{-(d+4+\Delta)+\frac{d+2+\Delta}{2} \tilde n +\frac{d-2+\Delta}{2} n}
\Gamma_{\tilde n,n}(\{(\ell^{1+\Delta} q_{\|}, \ell q_{\bot})\}, 
\ell^{4} \omega,
\ell^{2} \tau,\ell^{-2\sigma} u;\ell^{-1-\Delta}L_{\|},\ell^{-1} L_{\bot}),
\nonumber 
\end{eqnarray}
where $\Gamma_{\tilde n,n}$ is the one-particle irreducible correlation
function of $n$ density fields and $\tilde{n}$ response fields.
For the transverse structure factor it implies
\begin{equation}
\tilde{G}_\bot(q_\bot,\tau,u;,L_{\bot},L_{\|}) =
  \ell^2 \tilde{G}_\bot(\ell q_\bot,\ell^2 \tau, \ell^{-2\sigma}u;
     \ell^{-1} L_{\bot}, \ell^{-1-\Delta} L_{\|}).
\end{equation}
Setting $\ell = L_\bot$ and expanding for $L_\bot\to\infty$ we obtain for 
$2 < d < 5$
\begin{equation}
\tilde{G}_\bot(q_\bot,\tau,u;,L_{\bot},L_{\|}) = 
   L_\bot^2 f_2(q_\bot L_\bot, \tau L_\bot^2;S_\Delta) 
   [1 + O(uL^{-2\sigma}_\bot)].
\end{equation} 
Now the leading term is given by Eq.~\reff{sec4:tildeG}, and thus, in the 
absence of zero mode, which implies $|q_\bot| L_\bot \ge 2\pi$, 
$f_2(q_\bot L_\bot, \tau L_\bot^2;S_\Delta)$ is regular and finite in the 
whole high-temperature phase. Thus, the dangerously irrelevant coupling 
can be neglected for all $2 < d < 5$. 
Note that this argument  would not apply to the zero mode if it were present.
Indeed, for $q_\bot = 0$, $f_2(0, \tau L_\bot^2;S_\Delta) = 1/\tau L^2_\bot$ 
which is singular as $\tau \to 0$, giving rise to an anomalous behavior.
In App.~\ref{AppA} we discuss this phenomenon
in the large-$N$ limit of the $O(N)$ model above the upper critical dimension,
i.e. for $d > 4$, showing that no anomalous behavior is observed in the 
absence of zero mode (in particular the Binder parameter 
vanishes).\footnote{Even in the absence of zero mode it is still possible 
to observe anomalous scaling by performing a noncanonical scaling limit.
One should consider $\tau \to 0$, $L_\bot\to0$ with  
$\tau L_\bot^2 + 4 \pi^2 \to 0$ at the same time. 
See Ref. \cite{correlation_length} for a discussion 
in the $N$-vector model.} Therefore, for $d > 2$,
Eqs.~(\ref{eq:theoretical_pred}) and (\ref{eq:theoretical_pred_chi})
should hold without changes.
For $d =2$ the operator becomes marginal
($\sigma=0$) and therefore we expect logarithmic corrections to the
formulae previously computed. In the absence of any prediction, we will neglect
these logarithmic violations. As it has been observed in previous
numerical studies, if present, they are small \cite{Leung92}.
As we will discuss, this is confirmed by our numerical results.  

Finally, we wish to compute the behavior of the Binder parameter
keeping into account the presence of the dangerously irrelevant 
operator. Considering the connected static four-point correlation function 
for vanishing parallel momenta,
Eq.~(\ref{scaling-Gamma-nnt}) implies
\begin{equation}
\tilde{G}^{(4)} (\{(0,q_{\bot})\},\tau,u;L_{\|},L_{\bot}) = 
\ell^{d+4+\Delta} \tilde{G}^{(4)}   (\{(0,\ell q_{\bot})\}, 
\ell^{2} \tau,\ell^{-2\sigma} u;\ell^{-1-\Delta}L_{\|},\ell^{-1} L_{\bot}).
\end{equation}
Now $\tilde{G}^{(4)}$ is of order $u$, and therefore, setting $\ell = L_\bot$,
$q_\bot = \pm \bar{q}$ with $|\bar{q}| = 2\pi/L_\bot$ as in the definition of 
the Binder parameter, cf.~Eq.~\reff{defbinder}, we obtain
\begin{equation}
\tilde{G}^{(4)}
(\{(0,\pm \bar{q})\},\tau,u;L_{\|},L_{\bot}) = 
  u L_\bot^8 f_4(\tau L_\bot^2, S_\Delta)
   [1 + O(u L_\bot^{-2\sigma})].
\label{tildeG4-u}
\end{equation}
In App. \ref{AppB} we have verified this equation to first order 
in perturbation theory, proving also that $f_4(0,S_\Delta) \not = 0$.
It follows that the Binder parameter behaves as 
\begin{eqnarray}
g(\beta;L_{\|},L_{\bot}) &=& {u L_\bot^{5-d}\over L_\|} 
f_4(\tau L_\bot^2, S_\Delta) (4 \pi^2 + \tau L_\bot^2)^2 
     [1 + O(u L_\bot^{-2\sigma})]
\nonumber \\
&=& u L_\bot^{-2\sigma} f_g(\tau L_\bot^2, S_\Delta) 
     [1 + O(u L_\bot^{-2\sigma})], 
\end{eqnarray}
with $f_g(0,S_\Delta)\not = 0$. Therefore, for all $2 < d < 5$ the Binder 
parameter vanishes, in spite of the presence of the dangerously irrelevant
operator. In $d=2$ we expect logarithmic corrections. 
Note again that $g(\beta_c;L_{\|},L_{\bot})$ vanishes as $L_\bot\to \infty$ 
because $f_2(0,S_\Delta) \not= 0 $, a consequence of the fact that 
the Binder parameter we use here is defined at nonvanishing momenta.

Finally, we wish to discuss the distribution function of the order parameter
$\psi\equiv\phi(\ve{k}_{0,1})$ in the JSLC theory. 
For $\beta\to \beta_c$ and $L_\|$, 
$L_\bot\to\infty$, such a quantity has a Gaussian distribution, i.e. 
\begin{equation}
N \exp\left( - {|\psi|^2\over L_\| L_\bot^{d-1} \chi_\bot}\right) 
   d\psi d\psi^*,
\label{distribution-fun}
\end{equation}
where $N$ is a normalization factor,
and indeed the Binder parameter vanishes in this limit. Since 
the dangerously irrelevant operator is truly irrelevant, 
such a distribution is also valid at the critical point and in the FSS limit, 
allowing the computation of the distribution function of the magnetization.
We obtain
\begin{equation}
m^2 = {\pi\over 4} {\chi_\bot\over L_\| L_\bot^{d-1}}, 
\end{equation}
with corrections of order $L_\bot^{-2\sigma}$ 
(logarithms in $d=2$).
Then, we obtain
\begin{equation}
F_m(\alpha,z,S_\Delta) = {1\over \alpha^{(d+\Delta)/2}} 
      F_\chi(\alpha,z,S_\Delta)^{1/2},
\label{Fmtheor}
\end{equation}
with logarithmic corrections in $d=2$.

\section{Numerical Simulation}
\label{sec:simulations}

\subsection{Setup}

In order to 
study the critical behavior of the DLG in two dimensions, we perform an 
extensive Monte Carlo simulation. 
We use the dynamics described in Sec.~\ref{sec-2.1} with Metropolis rates,
i.e. we set
\begin{equation}
w(x) =\, {\rm min}\ (1,e^{-x}).
\end{equation}
Simulations are performed at infinite driving field. Therefore,
forward (backward) jumps in the direction of the field are always
accepted (rejected).

The dynamics of the \dlg\ is diffusive and the dynamic critical 
exponent associated with transverse fluctuations, which represent the slowest 
modes of the system, is expected to be close to 4 both in the JSLC and in the 
RDLG models. In the JSLC theory 
~\cite{Janssen86a} $z_\bot=4$ exactly . 
Thus, it is important to have
an efficient implementation of the Monte Carlo algorithm in
order to cope with  the severe critical slowing down.     

We use a multi-spin coding technique, evolving simultaneously $N_{\rm multi}$
independent configurations. We took particular care in optimizing 
the value $N_{\rm multi}$. On Pentium and PowerPC processors---the
computers we used in our simulations---we observed a nonmonotonic
behavior of the speed with $N_{\rm multi}$. 
For instance, on a Pentium processor,  we reach a speed of 
$1.3\cdot 10^8$ spin-flips/sec for $N_{\rm multi} = 32$ and
$2.7 \cdot 10^8$ for $N_{\rm multi} = 128$. The speed drops to
$2.0\cdot 10^8$ for $N_{\rm multi} = 192$ and then increases again,
reaching $3.8 \cdot 10^8$ for $N_{\rm multi} = 960$.
However, by increasing $N_{\rm multi}$ we increase the memory and 
disk requirements so that we used $N_{\rm multi} = 128$ as a good 
compromise.

For the pseudo-random numbers we use the Parisi-Rapuano congruential
generator, a 32 bit-shift-register generator based on 
\begin{equation}
a_n = (a_{n-24} + a_{n-55}) \,\text{XOR}\, a_{n-61},
\end{equation}
where $a_n$ is an unsigned 32-bit integer.

The main purpose of our work is to test the theoretical predictions of the
JSLC field-theoretical model
discussed in Sec.~\ref{sec-4}. It predicts $\Delta=2$ and therefore 
we have performed 
simulations on a set of lattices with $L_\|/L_\bot^3$ constant. 
We considered the following values of $(L_\|,L_\bot)$: $(21,14)$,
$(32,16)$, $(46,18)$, $(64,20)$, $(88,22)$, $(110,24)$,
$(168,28)$,$(216,30)$, $(262,32)$, $(373,36)$, $(512,40)$, $(592,42)$,
$(681,44)$, $(778,46)$,
$(884,48)$. It is easy to verify that $S_2 \approx 0.200$ (more precisely 
$0.197 \le S_2 \le 0.202$ for $L_\bot \le 28$ and 
$0.19992\le S_2 \le 1/5$ for $L_\bot \ge 30$). 
We considered several values of 
$\beta$: $0.28$, $0.29$, $0.3$, $0.305$, $0.3075$, $0.31$, $0.3105$,
$0.311$, $0.31125$, $0.3115$, $0.31175$ $0.312$. As we shall see below,
all lie in the
disordered phase, albeit very near to the critical point.
For a few values of $\beta$ we have also performed simulations 
for different values of $(L_\|,L_\bot)$. We have considered a sequence 
of lattices with $S_2\approx 0.300$ and $S_2\approx 0.100$ [the largest 
lattice has $(L_\|,L_\bot)$ = (1350,30), (414,48) respectively] and a sequence 
with $\Delta = 1$ and $S_1\approx 0.106$ [the largest
lattice has $(L_\|,L_\bot)$ = (245,48)]. The results for $\Delta = 2$ and 
$S_2 \approx 0.200$ are presented in Tables 
\ref{tab:observables1} and \ref{tab:observables2}.


It is very important to be sure that the system has reached the steady-state
distribution before sampling. 
Metastable configurations in which the system is trapped for times
much longer than typical relaxation times in the steady state are
a dangerous source of bias. In the \dlg ,
configurations with multiple stripes aligned with the external
field are very long-lived and may
persist for times of the order of a typical simulation run.
To avoid them, we started the simulations for  
the largest systems from suitably rescaled thermalized configurations
of smaller lattices at the same temperature and value of $S_\Delta$.  

We have performed a detailed study of the dynamic correlations. 
For $\beta = 0.312$, the value of $\beta$ that is nearest to $\beta_c$, 
we compute the autocorrelation time $\tau_\chi$ for the 
susceptibility $\chi_\bot$, which is expected to have a significant overlap 
with the slowest modes of the system. The results are reported 
in Table~\ref{Table-tauchi}. The autocorrelation times are expressed 
in sweeps, 
where a sweep is conventionally defined as the number of proposed
moves equal to the volume of the lattice. It is easy to verify that 
$\tau_\chi \sim L_\bot^4$ as expected.  For each $\beta$ and lattice size
we always make at least $3\cdot 10^7$ sweeps, so that our runs are at least,
but in most of the cases much longer than, approximately $10^3\, \tau_\chi$.

The statistical variance of the observables is estimated by using the 
jackknife method~\cite{Efron82}.   
To take into account the  correlations of the samples,
we used a blocking technique in the jackknife analysis, using blocks 
of $256\times 10^3$ sweeps. They are much longer than typical autocorrelation
times and thus different blocks are statistically independent.

\subsection{Results}
\label{sec5:FSS}

First, we check that the correlation length $\xi_\bot(\beta;L_\|,L_\bot)$ 
has a good thermodynamic limit, independent of the aspect ratio,  i.e. 
that, for $\beta<\beta_c$,  $\xi_\bot(\beta;L_\|,L_\bot)$ has a finite limit
when $L_\|,L_\bot\to \infty$ in an arbitrary way. 

Instead of $\xi_\bot(\beta;L_\|,L_\bot)$ we 
 consider (the reason of this definition will become clear
below)
\begin{equation}
  \label{eq:definition_of_tauelle}
  \tau(\beta; L_\|,L_\bot) \equiv 
  {1\over \xi_\bot^2(\beta;L_\|,L_\bot)} - {4 \pi^2 \over L_\bot^2}.
\end{equation}
Of course, if $\xi_\bot(\beta;L_\|,L_\bot)$ has a good thermodynamic limit, 
also $\tau(\beta; L_\|,L_\bot)$
has a finite limit and 
\begin{equation} 
\lim_{L_\|,L_\bot\to\infty} \tau(\beta; L_\|,L_\bot) = 
    {1\over \xi_{\bot,\infty}^2(\beta)}.
\label{limit-tauL}
\end{equation}

In Fig.~\ref{fig:tauelle1} we show $\tau(\beta;L_\|,L_\bot)$ for four values of 
$\beta$ and two different sequences of lattice sizes, one with 
$S_2\approx 0.200$ and the other one with $S_1 \approx 0.106$. The numerical 
results show that $\tau(\beta;L_\|,L_\bot)$ has a finite 
infinite-volume limit, which is reached from below 
using the data with fixed $S_2$ and from above using the data with fixed $S_1$.
The corrections to Eq.~(\ref{limit-tauL}) are not clear. 
In equilibrium systems the second-moment correlation length converges 
to $\xi_{\bot,\infty}(\beta)$ with corrections that decrease as $1/L_\|^2$, 
$1/L_\bot^2$, although in many cases, with an appropriate finite-volume 
definition, such corrections are so small that an apparent exponential 
convergence is observed \cite{CP-98}. 
These corrections can be easily related to the behavior of 
$\tilde{G}_{\bot,\infty}(q)$ in infinite volume. 
Indeed, $\tilde{G}_{\bot}(q;L_\|,L_\bot)$ converges to 
$\tilde{G}_{\bot,\infty}(q)$ exponentially and thus 
we obtain for $\xi_\bot \equiv \xi_{13}$ defined in Eq.~(\ref{defxi})
\begin{equation}
\xi_\bot^2(\beta;L_\|,L_\bot) = \xi_{\bot,\infty}^2\left[
1 + {\xi_{\bot,\infty}^2\over L_\bot^2} 
   \left(-4 \pi^2 + {10 \pi^2\over 3 \xi_{\bot,\infty}^2} + 
       {20 \pi^2 c(\beta)\over \xi_{\bot,\infty}^2} \right) + 
    O(L_\bot^{-4})\right],
\end{equation}
where 
\begin{equation}
c(\beta) = \left. {\partial^2 \tilde{G}^{-1}_{\bot,\infty}(q)/\partial (q^2)^2 
               \over 
                   \partial \tilde{G}^{-1}_{\bot,\infty}(q)^{-1}/\partial q^2} 
         \right|_{q=0}.
\end{equation}
This expression is valid for any $\beta$, not only near the critical point.
It shows that corrections vanish as $L_\bot^{-2}$,
although it does not allow to compute them, since $c(\beta)$, 
which for scaling reasons is expected to scale as $\xi_{\bot,\infty}^2$ as 
$\beta\to\beta_c$, is unknown. In the DLG case, we do not know what are 
the finite-volume corrections to the structure factor, but it is still 
reasonable to conjecture that corrections to $\xi_\bot$ 
vanish as $L_\bot^{-2}$. The results reported in Fig.~\ref{fig:tauelle1}
confirm this behavior. One apparently
observes a $L_\bot^{-2}$ correction, especially for the lattices with 
fixed $S_1$. 
It is interesting to observe that, considering the data with 
fixed $S_2$,  the finite-size corrections decrease as $\beta\to \beta_c$. 
This is due to our specific definition of $\tau(\beta;L_\|,L_\bot)$ 
and is the motivation for this choice. Indeed, Eq.~(\ref{fssxi2}) implies
\begin{equation}
\tau(\beta;L_\|,L_\bot) \approx {1\over \xi_{\bot,\infty}^2(\beta)} + 
   o(L_\bot^{-2}),
\label{tauL-correzioni}
\end{equation}
i.e. corrections vanish faster than $L_\bot^{-2}$, in the FSS limit 
at fixed $S_2$. The data in Fig.~\ref{fig:tauelle1} confirm this behavior
and thus support the JSLC prediction (\ref{fssxi2}).

For $\beta = 0.311$ we have performed a more detailed check by comparing 
results for three different values of $S_2$. We define
\begin{equation}
\delta(\beta;L_\|,L_\bot) \equiv \tau(\beta;L_\|,L_\bot) - \tau_\infty(\beta),
\end{equation}
where $\tau_\infty(\beta)$ is the extrapolated value of 
$\tau(\beta;L_\|,L_\bot)$ for $L_\bot=\infty$. 
If Eq.~(\ref{tauL-correzioni}) holds, 
such a quantity vanishes as $L_\bot^{-2-\omega}$ for $L_\bot\to\infty$, where 
$\omega>0$ is a correction-to-scaling exponent. 
In Fig.~\ref{fig:various-s2} we report $L_\bot^a \delta_L(\beta;L_\|,L_\bot)$ 
for three different values of $a$. For $a=2+\omega$, points with constant $S_2$
should lie on an approximately horizontal line. 
From Fig.~\ref{fig:various-s2} we obtain
$\omega \approx 1$, i.e. 
$\tau(\beta;L_\|,L_\bot) = \tau_\infty(\beta) + O(L_\bot^{-3})$ 
for $\beta$ close to $\beta_c$.

We wish now to perform a detailed FSS analysis comparing the numerical
data with the predictions (\ref{eq:theoretical_pred}) and 
(\ref{eq:theoretical_pred_chi}). First, we consider the correlation length 
using $\alpha = 2$. In Fig. ~\ref{fig:newfss2xi} 
we report our results for 
$\xi_\bot(\beta;8 L_\|,2 L_\bot)/\xi_\bot(\beta;L_\|,L_\bot)$ vs. 
$\xi_\bot(\beta;L_\|,L_\bot)/L_\bot$.
The solid line is the
theoretical prediction (\ref{eq:theoretical_pred}). 
It is clear that, as the size of the lattice
increases, the points converge towards the theoretical line. We would
like to emphasize that in this plot \emph{there are no tunable parameters}.
Thus, the observed collapse is very remarkable. To get rid of
the small corrections to FSS that are still present, in
Fig.~\ref{fig:newfss2xi2} we present the same data using on the 
horizontal axis $\xi_\bot(\beta;8 L_\|,2 L_\bot)/(2 L_\bot)$. 
By using the values of $\xi_\bot$ corresponding to the 
larger lattice, scaling corrections are systematically reduced.
Note that the very good agreement between theory and numerical data 
gives $\nu_\bot = 1/2$. Indeed, using 
Eq.~(\ref{eq:theoretical_pred}) and Eq.~(\ref{nu-da-F}) 
we obtain the mean-field value for the exponent $\nu_\bot$.

Next, we checked the FSS behavior of  the susceptibility.
In Fig.~\ref{fig:newfss2chi2} we report our numerical data together with 
the theoretical prediction (\ref{eq:theoretical_pred_chi}).
Again, we observe a very good agreement. This result immediately
implies $\gamma_\bot = 1$, cf. Eq.~(\ref{eq:dummy_dimensions}).

In Fig.~\ref{fig:newfss2mag2} we present the same plot for the 
magnetization. The results are in reasonable agreement with the 
theoretical prediction (\ref{Fmtheor}), although in the 
region of small $\xi(\beta;L_\|,L_\bot)/L_\bot$ one can see some deviations.
As we discuss below, such deviations can be interpreted as corrections to 
scaling, that in our case decay quite slowly, probably as an inverse power 
of $\log L_\bot$.  If we perform a polynomial fit of the data, we obtain 
$F_m(2,z^*,S_2) = 0.491(15)$ for $z^* = 1/(2 \pi)$. 
Using Eq.~(\ref{eq:dummy_dimensions}) we obtain 
\begin{equation}
    {\beta_\bot\over \nu_\bot} = 1.023(43).
\end{equation}
Such a result is in very good agreement with the JSLC prediction
$\beta_\bot/\nu_\bot = 1$. 

The results for the Binder parameter  reported in
Fig.~\ref{fig:newfss2binder2} show a good scaling behavior,
falling onto a single curve with small corrections.
By using Eq.~(\ref{eq:dummy_dimensions}) we obtain
approximately $g(\beta_c;L_\|,L_\bot) \sim L_\bot^{-0.45(15)}$.
Thus, $g=0$ at the critical point, in agreement with the JSLC theory 
that predicts Gaussian transverse fluctuations at the
critical point. The value of the exponent 
is difficult to interpret 
and we suspect that the Binder parameter is going to zero as 
some negative power of $\log L_\bot$, which however cannot be distinguished 
numerically from a small exponent.

Finally, we report 
results for the parallel correlation length. 
Fig.~\ref{fig:fssdelta} reports a plot of 
$\xi_\|(\beta;L_\|,L_\bot)/L_\|$ vs. $\xi_\bot(\beta;L_\|,L_\bot)/L_\bot$. 
There are very large corrections to FSS
but the data apparently collapse on a well-defined curve as the size of the 
lattice increases. This is again a very important test of the JSLC 
theory. Indeed, as discussed in Sec. \ref{sec:fss}
the data for the correlation length can both scale only if 
we have correctly chosen the anisotropy exponent $\Delta$. 
Thus, Fig.~\ref{fig:fssdelta}, obtained using data with $\Delta = 2$,
strongly supports such a value for the anisotropy exponent.

In the JSLC theory the order parameter has the Gaussian distribution 
(\ref{distribution-fun}). However, such an expression does not take 
into account the presence of the leading corrections and therefore 
the FSS behavior of the Binder parameter. If we include these 
terms we expect a more complex form 
\begin{equation}
   N\exp[- V(|\psi|^2)]\, d\psi d\psi^*.
\label{distr-general}
\end{equation}
For small  $|\psi|^2$ it is sensible to expand $V(|\psi|^2)$ in 
powers of $|\psi|^2$ and thus, if only the lowest moments of $|\psi|^2$
are of interest, we can try to approximate $V(|\psi|^2)$ with its 
first two terms, 
\begin{equation}
  V(|\psi|^2) \approx a |\psi|^2 + b |\psi|^4,
\label{V-general}
\end{equation}
where $b/a^2 \approx u L^{-2\sigma}_\bot$ times a function of $tL_\bot^2$. 

We now show numerically that this approximation describes remarkably well
our results for $m$, $\chi_\bot$, and $g$: 
Essentially all corrections to scaling 
we observe can be taken into account by simply assuming the 
distribution function (\ref{distr-general}) and (\ref{V-general}). 

If we define
\begin{eqnarray}
M_n(z) = {\int d\psi\, d\psi^* |\psi|^n e^{-|\psi|^2 - z |\psi|^4} \over 
          \int d\psi\, d\psi^* e^{-|\psi|^2 - z |\psi|^4} },
\end{eqnarray}
then the approximation (\ref{V-general}) predicts
\begin{eqnarray}
g &=& 2 - {M_4(z)\over M_2(z)^2} \nonumber \\
X &\equiv & L_\bot^{d-1} L_\| {m^2 \over \chi_\bot} = {M_1(z)^2\over M_2(z)},
\label{gX}
\end{eqnarray}
where $z = b/a^2$. Thus, $X$ turns out to be a function of 
the Binder parameter $g$, 
implicitly defined by Eq.~(\ref{gX}). 
In Fig.~\ref{fig:violets} we report $X$ versus $g$ together 
with the theoretical prediction (\ref{gX}). The agreement is remarkable,
indicating that all corrections are taken into account by a simple 
generalization of the distribution of the order parameter. 
Note that, as $g\to 0$, $X$ approaches $\pi/4\approx 0.785$, which is the value
expected for a purely Gaussian distribution.

Finally, we compute $\beta_c$ from the critical behavior of 
$\xi_{\bot,\infty}(\beta)$.
In order to compute the infinite-volume correlation length, 
we can use two strategies. 
One consists in extrapolating 
$\tau(\beta;L_\|,L_\bot)$ to $L_\bot\to\infty$ at fixed $S_2$. 
To minimize corrections to scaling we only use the data with $S_2\approx 0.200$.
In Table~\ref{tab:fit_tauelle} we show the results of the fits 
$\tau(\beta;L_\|,L_\bot) = \tau_\infty(\beta) + a(\beta) L_\bot^{-2}$ for 
each $\beta$. 
In all cases we performed several fits, including each time only the results
with $L\ge L_{\rm min}$ for increasing $L_{\rm min}$. 
The results we report correspond to the fit for which the 
$\chi^2$ is reasonable, i.e. it is less than the value corresponding 
to a 95\% confidence level. In two cases we have not been able to fulfil 
this criterion, with a $\chi^2$ which is however only slightly larger.
The value of $\tau_\infty(\beta)$ gives
the estimate of the infinite-volume correlation length
$\xi_{\bot,\infty}(\beta) = 1/\sqrt{\tau_\infty(\beta)}$. 

In the previous analysis we have not made any assumption. More 
precise estimates of $\xi_{\bot,\infty}(\beta)$ can be obtained if 
we assume the validity of the JSLC theory and 
use the method of 
Ref. \cite{Caracciolo95} to determine infinite-volume quantities.\footnote{
A similar method has been introduced in Refs. \cite{Luscher:1991wu,JKKim}.}
Such a method is particularly efficient and it has been successfully applied 
to many different equilibrium models 
\cite{Caracciolo95b,Caracciolo:1996ah,Mana-etal-96,Mendes-etal-96,Mana-etal-97,%
Ferreira-Sokal-99,Palassini-Caracciolo-99}. For the calculation 
we assume here the theoretical prediction for the function 
$F_\xi(\alpha,z,S_\Delta)$, cf. Eq~(\ref{eq:theoretical_pred}),
and use the iterative algorithm with 
$\alpha = 2$. 
To avoid scaling corrections we only consider the data 
with $L \ge L_{\text{min}} = 30$ and $S_2\approx 0.200$. 
Then, for each $\beta$ and $L_\bot$
we obtain estimates of $\xi_{\bot,\infty}(\beta)$. 
Results corresponding to the same $\beta$ are then 
combined to give the final estimate 
reported in Table \ref{tab:fss_extrapolation}. 
Here $N$ is the number of degrees of freedom 
(the number of data with the same $\beta$ minus one) and $R^2$ is the sum of 
square residuals which gives an indication of the consistency of the 
various extrapolations. The results are in agreement with those 
previously obtained, but are significantly more precise.

Once $\xi_{\bot,\infty}(\beta)$ has been computed,
we have first performed fits of the form
\begin{equation}
\xi_\infty(\beta) = \tilde A \left( 1-\frac{\beta}{\beta_c}\right)^{-\nu_\bot},
\label{fit-nubot}
\end{equation}
using only the data with
$\beta \ge \beta_{\text{min}}$. The value
$\beta_{\text{min}}$ corresponds to the lowest value for which
the $\chi^2$ is less than the value corresponding to 
a $95\%$ confidence level.  
Using the first set of extrapolated values,
see Table~\ref{tab:fit_tauelle}, we obtain 
$ \nu_\bot = 0.483(50)$, $\beta_c = 0.31264(15)$, 
while the second set, see Table~\ref{tab:fss_extrapolation}, gives 
$\nu_\bot = 0.500(32)$ and $\beta_c = 0.312696(88)$. The results
are in full agreement with the prediction $\nu_\bot=1/2$. 
Of course, in the second case this is simply a consistency check, 
since, by using the theoretical prediction for the scaling curve, 
we have implicitly assumed $\nu_\bot = 1/2$. 

In order to obtain our best estimate of $\beta_c$, we fix $\nu_\bot = 1/2$
and use the second set of extrapolations (which have been performed
implicitly assuming $\nu_\bot = 1/2$). 
Still fitting with Eq.~(\ref{fit-nubot}), we obtain for 
$\beta_{\text{min}}=0.31$
\begin{eqnarray}
 \beta_c & = 0.312694(18), \label{betacfinale}
\end{eqnarray}
with $\chi^2 = 6.0$ and $5$ degrees of freedom. 
The addition of 
an analytic correction does not improve substantially the result.

The result for $\beta_c$ should be compared with the existing determinations:
\begin{equation}
\beta_c = \begin{cases}
0.3108(11) & \text{(Ref.~\cite{Leung91b})}; \\
0.3125(13) & \text{(Ref.~\cite{Wang96})}; \\
0.3156(9) & \text{(Ref.~\cite{AGMM})}; \\
0.3125(10) & \text{(Ref.~\cite{AS-02})}. \\
\end{cases}
\end{equation}
Our result~(\ref{betacfinale}) and the estimates of 
Ref.~\cite{Leung91b,Wang96,AS-02} are in reasonable agreement. On the other 
hand, the estimate of Ref.~\cite{AGMM} is sligthly larger, 
although the difference (three error bars) is not yet very significant.

\section{Conclusions}
\label{sec:conclusions}

In this paper we have performed a thorough check of the theoretical predictions
for the DLG. We have verified that the FSS curves $F(2,z,S_\Delta)$ for 
$\chi_\bot$ and $\xi_\bot$ agree with the JSLC predictions 
(\ref{eq:theoretical_pred}) and (\ref{eq:theoretical_pred_chi}) if 
$\Delta$ is fixed to the JSLC value $\Delta=2$. We wish to stress 
that the comparison between theory and numerical data 
does not require any tuning of parameters, at variance with previous 
studies in which the FSS analysis required fixing 
$\beta_c$ and/or some exponents. We also analyzed the magnetization at 
the critical point finding $\beta_\bot/\nu_\bot = 1.023(43)$, 
in agreement with the JSLC value $\beta_\bot/\nu_\bot = 1$.
Our results for the Binder parameter $g$\ do not agree with those
of Ref. \cite{Leung92} where it was found
$g\not=0$ at criticality, but confirm 
the results of Wang \cite{Wang96} who could not find a satisfactory
collapse for the Binder parameter. Our result $g=0$ is compatible
with the idea that even finite-volume transverse correlations
are Gaussian in the scaling limit, so that $g=0$ at
criticality. Finally, we discuss the FSS behavior of the magnetization.
Our results are consistent with a Gaussian distribution for the 
order parameter with corrections that are well parametrized in
terms of the Binder parameter.

One point that is still unclear is the role of the dangerously irrelevant 
operator. A general analysis confirmed by an explicit one-loop computation
indicates that this operator does not give rise 
to FSS violations in any $2 < d < 5$. However, in two dimensions 
the same arguments predict logarithmic violations (if it is marginally
relevant) or at least logarithmic 
corrections (if it is marginally irrelevant). 
Numerically, we observe corrections to the Gaussian field-theoretical 
predictions but we are not able to determine their behavior as 
$L_\bot\to\infty$.
Indeed, very large lattices 
are needed to distinguish logarithmic corrections from power-law 
corrections with small exponents. In any case, our data indicate that 
the operator is marginally irrelevant since $g\to 0$ as $L_\bot \to 0$.
We find $\gamma_g/\nu_\bot = 0.45(15)$ but this result should not be taken
seriously. 
Probably, $g$ decreases as a power of $\log L$ because of the
marginal operator, but, in our range of values of $L$,
the complicated logarithmic dependence is mimicked by a single power.      
Note that, if $g(\beta_c;\infty,\infty) = 0$, the Binder parameter 
cannot be used to compute $\beta_c$: The crossing method does not work.

Let us now compare our results with those of Refs.~\cite{AGMM} and 
\cite{AS-02} that presented numerical results apparently in good 
agreement with the RDLG scenario. If the JSLC theory gives the correct 
description of the critical behavior, the family of lattices considered in 
Ref.~\cite{AGMM}---at fixed $\Delta=1$---is such that $L_\bot$ 
increases too fast compared to $L_\|$. As discussed in
Ref.~\cite{CGGP-shape}, in this case one expects to observe scaling 
with an effective geometry corresponding to $L_\bot = \infty$. Therefore,
longitudinal quantities should show the correct behavior, 
i.e. with the JSLC exponents, when studied in terms of 
$(\beta-\beta_c) L_\|^{1/\nu_\|}$. On the other hand,
transverse quantities should have 
a critical behavior with effective exponents.
Therefore, there is no contradiction with the estimates of Ref.~\cite{AGMM} 
that differ from the JSLC ones. 
It is also useful to take the opposite point of view: What should our results
be if the DLG belongs to the same universality class of the RDLG? If this 
were the case, the correct anisotropy exponent would be $\Delta = 1$. 
Thus, the lattices we consider are such that $L_\|$ increases too 
fast compared to $L_\bot$. Again, we expect to observe scaling 
\cite{CGGP-shape}, with an effective geometry $L_\| = \infty$. 
Transverse quantities should have the correct behavior, i.e. transverse 
critical exponents should coincide with those of the theory with $\Delta = 1$.
Therefore, our results should be the same as those of Ref.~\cite{AGMM}, which,
as we have shown here, is not the case. Thus, our data exclude 
$\Delta = 1$. As discussed in Ref.~\cite{CGGP-shape}, it is possible to 
determine $\Delta$ unambiguously from the FSS behavior of $\xi_\|$ and 
of $\xi_\bot$. Indeed, only if $\Delta$ is chosen correctly, both 
correlations length scale linearly, i.e. $\xi_\bot \sim L_\bot$ and 
$\xi_\|\sim L_\|$. This is indeed what we have checked in Sec.~\ref{sec5:FSS}.
Therefore, our data strongly support the JSLC prediction $\Delta = 2$, 
and do not support the claim of Ref.~\cite{AGMM} that the DLG and the 
RDLG belong to the same universality class.

Finally, let us discuss \cite{CGGP-commento} the results
of Albano and Saracco \cite{AS-02}. 
First of all, there is a serious flaw in one of their scaling Ans\"atze,
due to the fact that they do not distinguish between $z_\bot$ and $z_\|$
\cite{Schmittmann95}. While in their Eqs. (5), (6), and (7) the exponent
$z$ should be identified with $z_\|$ (the JSLC prediction is $z_\| = 4/3$
as correctly appears in their Table 1), in Eqs. (8), (9), and (10), 
the exponent $z$ should be identified with $z_\bot$ 
(the JSLC prediction is $z_\bot = 4$). Therefore, their result for 
$c_\perp$ does not agree with any 
prediction, neither the JSLC one nor the RDLG one. 
Moreover, with the lattice sizes they
consider, it is not clear whether they are really looking at the short-time 
dynamics  in infinite volume or rather to the approach to equilibrium 
in a finite lattice. Indeed, since the dynamics in the longitudinal direction
is fast, one expects to generate correlations of size $L_\|$ in a time of 
order $L_\|^{z_\|}$, so that a necessary condition to avoid size effect is 
that $t \ll L_\|^{z_\|}$. With their typical lattice sizes, one would expect to 
avoid size effects for $t \lesssim 10^2$. If larger values of $t$ are 
used---they consider times up to $10^4$---finite-size effects are 
relevant and thus, contrary to their claims, the aspect ratio and the 
value of the anisotropy exponent become again a crucial issue.

\bigskip\bigskip

It is a pleasure to thank Beate Schmittmann and Royce Zia for many 
useful discussions 
and Miguel Mu\~noz  for correspondence. We acknowledge the
support of the Istituto Nazionale Fisica della Materia (INFM) through 
the Advanced Parallel Computing Project at
CINECA. These computations were carried out in part on the Computational
Physics Cluster at New York University, which was supported by NSF grant
PHY--0116590.

\newcommand{\f}{\ensuremath \varphi}
\def\bsigma{\mbox{\protect\boldmath $\sigma$}}
\def\bmu{\mbox{\protect\boldmath $\mu$}}
\newcommand{\zm}{\mbox{{\sc zm}}}
\newcommand{\nzm}{\mbox{{\sc zm}\!\!\!\!\!\textbackslash\!\!\!\slash}\,\,}
\newcommand{\dd}{{\rm d}}

\appendix

\section{Binder cumulant without zero mode} \label{AppA}

In Sec.~\ref{sec-4} we stated that the Gaussian nature of the 
transverse fluctuations implies that the Binder cumulant vanishes
at criticality. We consider now the $O(N)$ model 
for $N\to\infty$ above the upper critical dimension ($d>4$), showing that the 
Binder cumulant defined in Eq.~\reff{defbinder} also 
vanishes, in spite of the presence of the 
dangerously irrelevant operator. This is due to the fact that the 
definition~\reff{defbinder} involves the correlation functions at 
nonvanishing momenta. 

To be concrete let us consider the $O(N)$ $\sigma$-model defined on a
finite hypercubic lattice $\Lambda$ of volume $L^d$, with Hamiltonian 
\begin{equation}
{\cal H} =\, - N \sum_{\langle\pos{x},\pos{y}\rangle}
\bsigma_\pos{x} \cdot \bsigma_\pos{y} ,
\end{equation}
where $\langle\cdot,\cdot\rangle$ indicates nearest-neighbor sites.
The partition function is simply
\begin{equation}
Z =\, \int \prod_\pos{x} [\dd\bsigma_\pos{x} \, 
\delta(\bsigma_\pos{x}^2-1)] \; e^{-\beta{\cal H}} \times
\left\{
\begin{array}{lc}
1 & \mbox{\zm} \\
\delta\left(\sum_{\pos{x}} \bsigma_\pos{x}\right)  & \mbox{\nzm} 
\end{array}
\right.
\end{equation}
where $\beta \equiv 1/T$, \zm\ corresponds to the standard case 
(with zero mode),
and \nzm\ to the theory without zero mode. 
The Fourier transform of the two-point correlation function is given by
\begin{equation}
G^{\alpha,\beta}(\ve{p}) \equiv \langle\sigma^\alpha_{-\ve{p}}
\sigma^\beta_{\ve{p}}\rangle_\Lambda =
\frac{1}{\beta}\frac{\delta_{\alpha,\beta}}{\hat{\ve{p}}^2 +
\lambda_L} \times
\left\{
\begin{array}{lc}
1 & \mbox{\zm} \\
1-\delta_{\ve{p},\ve{0}} & \mbox{\nzm} 
\end{array}
\right.
\end{equation}
where $\hat{\ve{p}}^2 = 4 \sum_\mu \sin^2(p_\mu/2)$
and $\lambda_L = \lambda_L(\beta,L)$ solves the finite-volume gap
equation of the model
\begin{equation}
\beta = {1\over L^d} \sum_{\pos{p}\in \Lambda^*}
\frac{1}{\hat{\ve{p}}^2 +
\lambda_L} \times
\left\{
\begin{array}{lc}
1 & \mbox{\zm} \\
1-\delta_{\ve{p},\ve{0}} & \mbox{\nzm} 
\end{array}
\right.
\end{equation}
in which
$\Lambda^* = {2\pi L^{-1}}\,\mathbb{Z}^d_{L}$
is the dual lattice. 

We now define two different Binder cumulants
\begin{eqnarray}
g_0(\beta,L) = - \frac{1}{L^d}
\frac{\langle(\ve{m}\cdot\ve{m})^2\rangle_c}{
      \langle\ve{m}\cdot\ve{m}\rangle^2_c}, \nonumber \\
g_1(\beta,L) = - \frac{1}{L^d}
\frac{\langle(\bmu\cdot\bmu)^2\rangle_c}{\langle\bmu\cdot\bmu\rangle^2_c} \, ,
\end{eqnarray}
where 
$m^\alpha\equiv\sigma^\alpha_{\ve{p}=\ve{0}}$,
$\mu^\alpha\equiv\sigma^\alpha_{\ve{p}=\ve{p}_{\rm min}}$, 
and $|\ve{p}_{\rm min}| = 2\pi/L$. The definition $g_0(\beta,L)$ is the 
one usually used in the $N$-vector model, but it is unsuitable
for systems without zero mode. In this case, the natural definition is 
$g_1(\beta,L)$ and indeed such a quantity corresponds to the 
Binder cumulant we have used in the DLG simulation, cf. Eq.~\reff{defbinder}.

Now, let us compute $g_{0,1}(\beta_c,L)$ for $d>4$ at the critical point 
$\beta_c$, 
\begin{equation}
\beta_c =
\int_{-\pi}^{\pi}\frac{\dd^dq}{(2\pi)^d}\frac{1}{\hat{\ve{q}}^2} \, .
\end{equation}
In Ref.~\cite{correlation_length} we found that, for $d>4$,
\begin{equation}
\lambda_L(\beta_c,L) =  
\left\{
\begin{array}{lc}
{\cal C}_{d,1}^{-1/2} L^{-d/2}[1 + O(L^{2-d/2})] & \zm \\
{\cal I}(0)/{\cal C}_{d,1} L^{2-d}[1+O(L^{4-d})] & \nzm
\end{array}
\right.
\end{equation}
where ${\cal C}_{d,1}$ and ${\cal I}(\rho)$ are defined in Eqs.~(2.19) and
(A.5) of Ref.~\cite{correlation_length}, respectively. This is
sufficient to determine the behavior of $\langle
\ve{m}\cdot\ve{m}\rangle_c$ 
and $\langle\bmu\cdot\bmu\rangle_c$ at (bulk) criticality. As far as
the four-point function is concerned, it can be expressed in terms of
\begin{equation}
\Delta^{-1}(\ve{q}) \equiv \frac{1}{2\beta^2} 
{1\over L^d} \sum_{\pos{p}\in \Lambda^*}
\frac{1}{[\widehat{\ve{p} + \ve{q}}^2 +
\lambda_L][\widehat{\ve{p}}^2 +
\lambda_L]}  \times 
\left\{
\begin{array}{lc}
1 & \mbox{\zm} \\
(1-\delta_{\ve{p},\ve{0}})(1-\delta_{\ve{p}+\ve{q},\ve{0}}) & \mbox{\nzm} 
\end{array}
\right.
\end{equation}
After some calculation one can easily show that, at criticality and
with $|\ve{q}|\sim 1/L$,
\begin{equation}
\left. \Delta^{-1}(\ve{q})\right|_{\beta_c} = \frac{{\cal C}_{d,1}}{2\beta_c^2}
(1+\delta_{\ve{q},\ve{0}}) + 
\left\{
\begin{array}{lc}
O(L^{2-d/2})& \mbox{\zm} \\
O(L^{4-d}) & \mbox{\nzm} 
\end{array}
\right.
\end{equation}
In the large $N$-limit we have
\begin{align}
\langle \bsigma_{-\ve{q}}\cdot\bsigma_{\ve{q}}\rangle_c &=
N G^{1,1}(\ve{q}) + O(N^0), \\
\langle (\bsigma_{-\ve{q}}\cdot\bsigma_{\ve{q}})^2\rangle_c &= - 
(N + 2) \Delta(\ve{q}) [G^{1,1}(\ve{q})]^4 + O(N^0), 
\end{align}
so that 
\begin{align}
g_0(\beta_c,L) &= \frac{N+2}{N^2} [1+O(L^{2-d/2},N^{-2})] ,&\quad \zm\\
g_1(\beta_c,L) &= O(N^{-1}L^{4-d}) . &\quad \zm,\ \nzm
\end{align}
The different scaling behavior can be traced back to the fact that,
while the four-point function has always the same scaling behavior,
irrespective of the momentum at which it is computed, the two-point
function scales differently (for $d>4$) depending on whether it is
computed at zero or at nonzero momentum.

It is also of interest to compute the Binder parameter in $d=4$. 
Without reporting the details, we find that $g_0(\beta_c,L)$ 
agrees with the result reported above, with logarithmic corrections. 
For $g_1(\beta_c,L)$ we obtain instead
\begin{equation}
  g_1(\beta,L) \approx {N+2\over N^2} {1\over \pi^2 \log L}.
\end{equation}
The Binder cumulant $g_1(\beta,L)$ vanishes also in $d=4$, albeit only
logarithmically.

\section{Dangerously irrelevant operator and correlation functions} 
\label{AppB}

In this Appendix we compute the zero-momentum insertion of the dangerously
irrelevant operator ${\cal A}$ into static correlation functions of 
$n$ density fields $s$ taken at vanishing parallel momenta 
in the JSLC theory \cite{Janssen86a}.\footnote{In this Appendix we use the 
notations of Ref.~\cite{Janssen86a}.}
We begin by considering 
$\f(\ve{p},\omega) \equiv s(p_\| =0,\ve{p}_\bot = \ve{p},\omega)$, 
i.e. we compute
\begin{equation}
\langle \f(\ve{p}_1,\omega_1) \cdots \f(\ve{p}_n,\omega_n) {\cal A}
   \rangle_{\rm conn} .
\label{insertion}
\end{equation}
We recall that
the interaction vertex ${\cal V}$ of the theory \cite{Janssen86a} is 
$\ts \grpa s^2$  ($\ts$ is the response field)
and thus it vanishes whenever the
parallel momenta flowing into it from the $\ts$-leg is zero. 
Let us discuss the consequences of this fact on the form of the
generic diagram ${\cal D}$ contributing to Eq.~\reff{insertion}.  

First, note that if the amputated diagram ${\cal D}^{\rm amp}$
associated with ${\cal D}$ has no external $\ts$-legs, it contains a loop
of response propagators and therefore it
vanishes because of causality \cite{Janssen76,Bausch76,Janssen79}.
Thus ${\cal D}$ vanishes unless ${\cal D}^{\rm amp}$ has at least one
external $\ts$-leg. 
Each external $\ts$-leg in ${\cal D}^{\rm amp}$ 
is connected either to 
the insertion ${\cal A}$ or to an interaction vertex 
${\cal V}$. In the latter case, however,
the contribution vanishes because of the zero parallel momentum
flowing from the external $\ts$-leg into the vertex ${\cal V}$. 
Thus, 
${\cal D}$ does not vanish only if each of the $\tilde{n}\ge1$ 
external $\ts$-legs of ${\cal D}^{\rm amp}$ belongs
to the insertion of ${\cal A}$. 
Now we show that ${\cal D}^{\rm amp}$
cannot be one-particle reducible. 
Indeed, let us suppose that ${\cal D}^{\rm amp}$ 
can be divided into two amputated subdiagrams ${\cal D}^{\rm amp}_1$ 
and ${\cal D}^{\rm amp}_2$, such that ${\cal A}\in{\cal D}^{\rm amp}_1$,
connected with a single line. 
By momentum conservation, the parallel momentum flowing into
this line is zero and therefore such a line must be connected to an $s$-leg
of a vertex in ${\cal D}_2^{\rm amp}$.
Since we also have ${\cal A}\in{\cal D}^{\rm amp}_1$, 
all external legs of ${\cal D}^{\rm amp}_2$ are of $s$-type. 
By causality ${\cal D}^{\rm amp}_2$ vanishes.

Summing up, the amputated diagrams contributing to
Eq.~\reff{insertion} are one-particle irreducible and all the $\tilde{n}\ge
1$ external $\ts$-legs belong to the insertion of ${\cal A}$.

This conclusion is independent of the particular ${\cal A}$
considered. Now let us specify our discussion to the case of the
dangerously irrelevant operator. The dangerous irrelevant coupling 
$\ts\nbpe s^3$ mixes, under renormalization-group transformations, 
with other operators of equal or
smaller scaling dimension. Those with the same dimension, that we
consider in the following, are listed in 
Ref.~\cite{Janssen86a}. 
They have the following general structure (the dots 
indicate derivatives): ($a$) $\ts (\cdots) s$, ($b$)
$\ts (\cdots) \ts$, ($c$) $\ts \grpa (\cdots) s^2$,
($d$) $\ts (\cdots) s^3$.
For $n > 2$, the only case we consider,
the insertion of one of the operators ($a$) or ($b$) into
the correlation function gives only
one-particle reducible contributions since one of their $\ts$-legs
has to be an external one for the amputated diagram.\footnote{For $n=2$
($a$) and ($b$) have nonzero tree-level contributions.} Thus, according
to our previous discussion, they do not 
contribute to the correlation function. The same is true for operators 
($c$). The $\ts$-leg, 
being an external one for the amputated diagram,
carries a zero parallel momentum $p_\|$ into the vertex. Since
these vertices are proportional
to  $p_\|$, they vanish.
Thus, contributions can only come from $\ts \nbpe s^3\in {\cal A}$. 
Defining
\begin{equation}
B_3 \equiv \frac{1}{6}\int \dd t \dd^dx \, \ts \nbpe s^3 ,
\end{equation}
we obtain 
\begin{equation}
\langle \f(\ve{p}_1,\omega_1) \cdots \f(\ve{p}_n,\omega_n) {\cal A}
   \rangle_{\rm conn,amp}  =\, 
u \Gamma_{1,n-1;B_3} , 
\label{final-expr}
\end{equation}
where $\Gamma_{\tilde{n},n;B_3}$ is the one-particle irreducible
vertex function with $\tilde{n}$ external $\ts$-legs, $n$ external
$s$-legs, and one (zero-momentum) insertion of the operator $B_3$,
${\cal A} \approx u B_3 + \ldots$ Therefore, all connected correlation 
functions with vanishing parallel momenta and $n>2$ are proportional to $u$ 
as $u\to 0$.

We wish now to compute the transverse static four-point function
$\tilde{G}^{(4)}(\ve{p}, \ve{p}, -\ve{p}, -\ve{p}; L_\|,L_\bot) $
at zero parallel momenta. For simplicity, we restrict the computation to 
$\tau = 0$, i.e. to the critical point.
The tree-level insertion of $B_3$ into the four-point
correlation functions with all fields $s$ taken at the same value of
$t$ (we can set $t=0$) and with momenta $\ve{P} = (0,\ve{p}_\perp)$ is given by
\begin{equation}
{\mathcal D}_{\rm TL} = 4 u_3 \ve{p}^2_\perp \int_{-\infty}^{+\infty}\!\dd
t_i\, R(-t_i,\ve{P})C^3(-t_i,\ve{P}) = \frac{u_3}{(\ve{p}_\perp^2)^4},
\end{equation}
where we have used\footnote{We disregard dimensionless factors present in 
the dynamic functional of Ref.~\cite{Janssen86a}.
}
(see Ref.~\cite{Janssen86a}) 
\begin{align}
R(t_1-t_2,\ve{q})&=\langle\ts(t_2,-\ve{q}) s(t_1,\ve{q})\rangle_{\rm
 TL} \nonumber\\
&=\theta(t_1-t_2) \exp\{-[\ve{q}_\bot^4+q_\|^2](t_1-t_2)\}\\
C(t_1-t_2,\ve{q})&=\langle s(t_2,-\ve{q}) s(t_1,\ve{q})\rangle_{\rm
 TL} \nonumber\\
& =\frac{\ve{q}_\bot^2}{\ve{q}_\bot^4+q_\|^2}
    \exp\{-[\ve{q}_\bot^4+q_\|^2]|t_1-t_2|\},
\end{align}
and $\theta(t)$ is the Heaviside step function.

At one loop there are two diagrams whose contributions can be written in the 
form
\begin{align}
&L_1(t_1,t_2,t_3;\ve{p}_\perp) = \nonumber \\
&\quad - g^2 \frac{1}{V}\sum_{(q_\|,\ve{q}_\perp)}\!\!\!\!\!'\ \ve{q}^{1,2}_\| \ve{q}^{1,3}_\|
R(t_1-t_2,\ve{q}^{1,2}) R(t_1-t_3,\ve{q}^{1,3}) C(t_2-t_3,\ve{q}^{2,3}),\\
&L_2(t_1,t_2,t_3;\ve{p}_\perp) = \nonumber \\
&\quad - g^2 \frac{1}{V}\sum_{(q_\|,\ve{q}_\perp)}\!\!\!\!\!' \ \ve{q}^{1,2}_\| \ve{q}^{2,3}_\|
R(t_1-t_2,\ve{q}^{1,2}) R(t_2-t_3,\ve{q}^{2,3}) C(t_1-t_3,\ve{q}^{1,3}),
\end{align}
where $\sum'$ runs over the whole momentum space {\it except} the zero
mode $(0,\ve{0})$ {\it and} $(0,\pm\ve{p}_\perp)$, and 
\begin{align}
&\ve{q}^{1,2} = (q_\|, \ve{q}_\perp + \ve{p}_\perp), \\ 
&\ve{q}^{1,3} = (-q_\|, -\ve{q}_\perp + \ve{p}_\perp),\\
&\ve{q}^{2,3} = (q_\|, \ve{q}_\perp).
\end{align}
The final result is given by $T = T_1+T_2$ where we defined
\begin{equation}
T_i(\ve{p}_\perp) = 
\gamma_i u_3 \ve{p}^2_\perp\int_{-\infty}^{+\infty}\!\!\dd t_1\dd
t_2 \dd t_3 
L_i(t_1,t_2,t_3;\ve{p}_\perp) R(-t_1,\ve{P})\prod_{k=1}^3 C(-t_k,\ve{P}),
\end{equation}
and $\gamma_1 = 4!/2$, $\gamma_2 = 2\gamma_1$ are combinatorial factors.
Performing the integrations over $t_i$, it is easy to find
\begin{equation}
T(\ve{p}_\perp) = \frac{\gamma_1}{2}g^2\frac{u_3}{(\ve{p}_\perp^2)^3}
 \frac{1}{V}\sum_{(q_\|,\ve{q}_\perp)}\!\!\!\!' \ f(\ve{q},\ve{p}_\perp),
\end{equation}
with $f(\ve{q},\ve{p}_\perp = \ve{0}) \not= 0$ and $f((\lambda^2 q_\|,
\lambda \ve{q}_\perp),\ve{p}_\perp)\sim \lambda^{-8}$ for large
$\lambda$. 
The sum over 
$f(\ve{q},\ve{p}_\perp)$ is convergent by power counting (for $d < 7$)
and thus, for $|p_\bot| \to 0$, 
\begin{equation}
\sum_{(q_\|,\ve{q}_\perp)}\!\!\!\!' \ f(\ve{q},\ve{p}_\perp) 
\approx \sum_{(q_\|,\ve{q}_\perp)}\!\!\!\!' \ f(\ve{q},0) \sim L_\bot^8.
\end{equation}
Then, if $|\ve{p}_\perp|\sim 1/L_\perp$, it is easy to see that, for
$d=5$ and at $S = L_\perp^2/L_\|$ fixed
(thus $V = L_\perp^{d+1}/S$), $T$ is finite and
\begin{equation}
T \sim {\cal D}_{\rm TL} \sim L_\perp^8
\end{equation}
in agreement with Eq.~(\ref{tildeG4-u}).


\newpage

\begin{table}
\tiny  \centering

\begin{tabular}{llllllll}
 \hline 
 \hfil $L_{\|}$ \hfil & \hfil $L_\bot$ \hfil &  \hfil $\xi_\bot$ \hfil & \hfil 
  $\chi_\bot$ \hfil & \hfil $A$ \hfil & \hfil $m$ \hfil & \hfil $g$ \hfil & \hfil 
  $\xi_{\|}$ \hfil  \\ 
 \hline 
 & & \multicolumn{6}{c}{$ \beta = 0.28 $} \\ 
 \hline 
32 & 16 & 1.6191(18) & 10.758(12) & 0.24368(40) & 0.130291(73) & 0.19471(80) & 1.5953(65)\\ 
46 & 18 & 1.6939(19) & 11.782(13) & 0.24353(42) & 0.106822(60) & 0.14899(76) & 2.5022(91)\\ 
64 & 20 & 1.7486(24) & 12.577(15) & 0.24312(52) & 0.088528(55) & 0.11355(89) & 3.641(10)\\ 
88 & 22 & 1.7952(23) & 13.221(16) & 0.24375(45) & 0.073648(45) & 0.08358(94) & 5.122(12)\\ 
110 & 24 & 1.8295(27) & 13.731(17) & 0.24375(55) & 0.064195(40) & 0.0663(11) & 6.346(13)\\ 
168 & 28 & 1.8838(33) & 14.568(22) & 0.24359(60) & 0.049452(37) & 0.0416(11) & 9.453(23)\\ 
262 & 32 & 1.9130(38) & 15.109(23) & 0.24221(70) & 0.037688(30) & 0.0277(11) & 13.582(41)\\ 
373 & 36 & 1.9339(50) & 15.484(28) & 0.24155(93) & 0.030126(27) & 0.0167(13) & 17.804(73)\\ 
512 & 40 & 1.9589(64) & 15.839(38) & 0.2423(12) & 0.024660(29) & 0.0089(24) & 22.22(13)\\ 
\hline 
  & & \multicolumn{6}{c}{$ \beta = 0.29 $} \\ 
 \hline 
32 & 16 & 1.8263(17) & 13.231(13) & 0.25208(39) & 0.145232(73) & 0.25367(72) & 1.7096(61)\\ 
46 & 18 & 1.9385(20) & 14.892(18) & 0.25235(37) & 0.120651(76) & 0.20415(84) & 2.6886(74)\\ 
64 & 20 & 2.0284(26) & 16.334(19) & 0.25189(49) & 0.101281(61) & 0.16350(80) & 3.9684(87)\\ 
88 & 22 & 2.1010(25) & 17.559(23) & 0.25139(41) & 0.085143(56) & 0.12623(83) & 5.625(10)\\ 
110 & 24 & 2.1626(26) & 18.572(24) & 0.25184(41) & 0.074864(49) & 0.10536(88) & 7.064(12)\\ 
168 & 28 & 2.2486(38) & 20.138(33) & 0.25108(56) & 0.058257(49) & 0.0701(11) & 10.627(18)\\ 
262 & 32 & 2.3090(40) & 21.315(37) & 0.25011(56) & 0.044811(40) & 0.0437(13) & 15.740(35)\\ 
373 & 36 & 2.3518(49) & 22.110(44) & 0.25016(66) & 0.036028(36) & 0.0285(14) & 21.017(54)\\ 
512 & 40 & 2.387(14) & 22.84(10) & 0.2494(21) & 0.029636(61) & 0.0165(55) & 27.09(49)\\ 
\hline 
  & & \multicolumn{6}{c}{$ \beta = 0.30 $} \\ 
 \hline 
32 & 16 & 2.0810(19) & 16.654(18) & 0.26003(20) & 0.16414(12) & 0.3287(11) & 1.8128(99)\\ 
46 & 18 & 2.2599(26) & 19.612(20) & 0.26040(55) & 0.139479(84) & 0.2857(11) & 2.8963(90)\\ 
64 & 20 & 2.4111(19) & 22.351(29) & 0.26009(50) & 0.119294(98) & 0.2426(13) & 4.3016(32)\\ 
88 & 22 & 2.5570(29) & 25.083(30) & 0.26066(45) & 0.102430(75) & 0.20601(64) & 6.202(16)\\ 
110 & 24 & 2.6604(17) & 27.218(18) & 0.26004(21) & 0.091161(32) & 0.17936(56) & 7.8399(71)\\ 
168 & 28 & 2.8383(41) & 31.017(47) & 0.25972(43) & 0.072636(58) & 0.13168(89) & 11.964(14)\\ 
262 & 32 & 2.9828(34) & 34.254(60) & 0.25974(33) & 0.057022(57) & 0.0952(16) & 18.223(30)\\ 
373 & 36 & 3.0875(46) & 36.748(72) & 0.25942(34) & 0.046562(47) & 0.0643(12) & 24.984(81)\\ 
512 & 40 & 3.1673(69) & 38.735(99) & 0.25899(53) & 0.038667(45) & 0.0487(26) & 32.511(60)\\ 
\hline 
  & & \multicolumn{6}{c}{$ \beta = 0.3025 $} \\ 
 \hline 
32 & 16 & 2.1529(18) & 17.656(16) & 0.26252(34) & 0.169306(83) & 0.34838(66) & 1.8477(56)\\ 
46 & 18 & 2.3538(21) & 21.087(22) & 0.26274(33) & 0.144925(78) & 0.30826(63) & 2.9442(74)\\ 
64 & 20 & 2.5344(27) & 24.444(31) & 0.26276(35) & 0.125037(84) & 0.26921(72) & 4.4080(77)\\ 
88 & 22 & 2.6955(29) & 27.681(36) & 0.26249(38) & 0.107824(73) & 0.23164(74) & 6.3251(94)\\ 
110 & 24 & 2.8283(30) & 30.490(40) & 0.26235(40) & 0.096697(67) & 0.20661(81) & 8.033(11)\\ 
168 & 28 & 3.0548(37) & 35.567(58) & 0.26238(41) & 0.077917(66) & 0.1559(11) & 12.367(16)\\ 
262 & 32 & 3.2340(55) & 39.948(89) & 0.26181(44) & 0.061649(71) & 0.1127(14) & 18.913(30)\\ 
373 & 36 & 3.3720(64) & 43.30(11) & 0.26258(45) & 0.050597(68) & 0.0819(16) & 26.068(50)\\ 
512 & 40 & 3.4664(82) & 45.92(14) & 0.26170(58) & 0.042138(63) & 0.0598(23) & 34.334(90)\\ 
\hline 
  & & \multicolumn{6}{c}{$ \beta = 0.3050 $} \\ 
 \hline 
32 & 16 & 2.2299(16) & 18.781(16) & 0.26476(24) & 0.175010(84) &  &   \\ 
46 & 18 & 2.4605(22) & 22.815(26) & 0.26534(26) & 0.151172(96) &   &   \\ 
64 & 20 & 2.6633(27) & 26.749(38) & 0.26517(27) & 0.13114(10) &    &   \\ 
88 & 22 & 2.8588(35) & 30.793(54) & 0.26541(29) & 0.11403(11) &    &   \\ 
110 & 24 & 3.0259(43) & 34.462(70) & 0.26570(29) & 0.10308(11) &    &   \\ 
168 & 28 & 3.3077(62) & 41.22(11) & 0.26543(36) & 0.08412(12) &    &   \\ 
216 & 30 & 3.4114(70) & 44.06(13) & 0.26415(36) & 0.07390(12) &    &   \\ 
256 & 32 & 3.5286(85) & 47.07(17) & 0.26450(42) & 0.06786(13) &    &   \\ 
365 & 36 & 3.710(11) & 52.05(22) & 0.26442(50) & 0.05621(12) &    &   \\ 
500 & 40 & 3.888(14) & 56.95(29) & 0.26548(64) & 0.04759(13) &    &   \\ 
592 & 42 & 3.914(16) & 57.93(31) & 0.26443(99) & 0.04298(12) & 0.0714(39) & 41.03(19)\\ 
681 & 44 & 3.979(15) & 60.18(30) & 0.26308(84) & 0.039892(98) & 0.0653(30) & 46.27(16)\\ 
778 & 46 & 4.038(10) & 61.62(20) & 0.26462(57) & 0.036916(61) & 0.0589(20) & 51.86(14)\\ 
884 & 48 & 4.0554(99) & 62.51(19) & 0.26310(58) & 0.034121(55) & 0.0483(24) & 57.63(16)\\ 
\hline 
  & & \multicolumn{6}{c}{$ \beta = 0.3075 $} \\ 
 \hline 
32 & 16 & 2.3093(21) & 19.992(18) & 0.26676(35) & 0.181034(88) & 0.39515(57) & 1.9069(49)\\ 
46 & 18 & 2.5687(19) & 24.682(23) & 0.26733(32) & 0.157657(78) & 0.36309(69) & 3.0624(65)\\ 
64 & 20 & 2.8141(24) & 29.501(31) & 0.26844(33) & 0.138181(79) & 0.33179(70) & 4.5961(80)\\ 
88 & 22 & 3.0400(30) & 34.533(43) & 0.26762(32) & 0.121188(79) & 0.29934(68) & 6.6479(79)\\ 
168 & 28 & 3.6017(45) & 48.397(93) & 0.26804(31) & 0.091505(93) & 0.2356(11) & 13.145(14)\\ 
262 & 32 & 3.9228(68) & 57.37(14) & 0.26825(43) & 0.074304(94) & 0.1859(14) & 20.397(27)\\ 
373 & 36 & 4.1764(87) & 65.23(18) & 0.26740(45) & 0.062411(92) & 0.1485(16) & 28.500(45)\\ 
512 & 40 & 4.374(10) & 71.66(24) & 0.26701(48) & 0.052849(94) & 0.1167(24) & 38.117(78)\\ 
681 & 44 & 4.549(14) & 77.60(34) & 0.26669(54) & 0.04540(10) & 0.0950(21) & 48.88(12)\\ 
\hline 
 \end{tabular}
  \caption{The results of our simulations 
  with $\Delta = 2$ and $S_2\approx 0.200$. 
  Here $A \equiv \xi^2_\bot/\chi_\bot$. 
 (table continues as Table~\ref{tab:observables2}).}
  \label{tab:observables1}
\end{table}

\begin{table}
\tiny  \centering

\begin{tabular}{llllllll}
 \hline 
 \hfil $L_{\|}$ \hfil & \hfil $L_\bot$ \hfil &  \hfil $\xi_\bot$ \hfil & 
 \hfil $\chi_\bot$ \hfil & \hfil $A$ \hfil & \hfil $m$ \hfil & 
 \hfil $g$ \hfil & \hfil $\xi_{\|}$ \hfil  \\ 
\hline 
  & & \multicolumn{6}{c}{$ \beta = 0.31 $} \\ 
 \hline 
32 & 16 & 2.3975(28) & 21.392(31) & 0.26870(25) & 0.18784(15) & 0.42136(99) & 1.9320(82)\\ 
46 & 18 & 2.6893(22) & 26.803(21) & 0.26983(48) & 0.164882(77) & 0.39503(46) & 3.1211(84)\\ 
64 & 20 & 2.9691(13) & 32.596(27) & 0.27044(27) & 0.145826(79) & 0.36797(77) & 4.6893(89)\\ 
88 & 22 & 3.2437(43) & 39.005(63) & 0.26975(63) & 0.12933(12) & 0.3399(14) & 6.7983(51)\\ 
110 & 24 & 3.4866(24) & 44.941(38) & 0.27049(21) & 0.118671(53) & 0.32316(66) & 8.6928(80)\\ 
168 & 28 & 3.9556(44) & 57.703(84) & 0.27116(28) & 0.100412(81) & 0.2893(11) & 13.564(13)\\ 
262 & 32 & 4.3976(58) & 71.31(12) & 0.27119(33) & 0.083257(82) & 0.2425(15) & 21.143(30)\\ 
373 & 36 & 4.7836(81) & 84.43(20) & 0.27101(36) & 0.071388(92) & 0.2109(18) & 29.947(46)\\ 
512 & 40 & 5.112(12) & 96.54(34) & 0.27073(41) & 0.06163(12) & 0.1760(23) & 40.289(97)\\ 
\hline 
  & & \multicolumn{6}{c}{$ \beta = 0.3105 $} \\ 
 \hline 
32 & 16 & 2.4141(18) & 21.584(17) & 0.27000(31) & 0.188709(78) & 0.42424(50) & 1.9437(49)\\ 
46 & 18 & 2.7082(23) & 27.167(24) & 0.26996(34) & 0.166028(80) & 0.39866(62) & 3.1332(61)\\ 
64 & 20 & 3.0040(30) & 33.295(35) & 0.27104(40) & 0.147436(82) & 0.37382(59) & 4.7005(70)\\ 
88 & 22 & 3.2910(33) & 39.952(49) & 0.27109(35) & 0.130987(88) & 0.34836(80) & 6.8253(92)\\ 
110 & 24 & 3.5446(38) & 46.274(61) & 0.27151(37) & 0.120523(85) & 0.33253(72) & 8.749(11)\\ 
168 & 28 & 4.0443(58) & 60.03(12) & 0.27247(37) & 0.10250(11) & 0.2993(12) & 13.658(15)\\ 
262 & 32 & 4.5097(78) & 74.75(19) & 0.27205(37) & 0.08534(12) & 0.2559(13) & 21.358(23)\\ 
373 & 36 & 4.940(11) & 89.79(30) & 0.27177(41) & 0.07372(13) & 0.2281(18) & 30.152(49)\\ 
512 & 40 & 5.291(13) & 103.04(38) & 0.27166(45) & 0.06374(13) & 0.1934(20) & 40.584(81)\\ 
\hline 
  & & \multicolumn{6}{c}{$ \beta = 0.3110 $} \\ 
 \hline 
32 & 16 & 2.4295(20) & 21.876(18) & 0.26982(35) & 0.190095(85) & 0.42942(53) & 1.9526(50)\\ 
46 & 18 & 2.7330(23) & 27.666(25) & 0.26998(34) & 0.167675(85) & 0.40611(63) & 3.1384(68)\\ 
64 & 20 & 3.0375(31) & 34.004(36) & 0.27133(39) & 0.149149(86) & 0.38293(61) & 4.7344(75)\\ 
88 & 22 & 3.3367(30) & 40.949(49) & 0.27189(36) & 0.132722(85) & 0.35717(70) & 6.8476(85)\\ 
110 & 24 & 3.6087(35) & 47.810(61) & 0.27239(36) & 0.122655(85) & 0.34509(79) & 8.787(10)\\ 
168 & 28 & 4.1255(52) & 62.29(11) & 0.27324(33) & 0.10453(10) & 0.3117(11) & 13.767(15)\\ 
262 & 32 & 4.6328(72) & 78.78(19) & 0.27243(35) & 0.08778(11) & 0.2775(12) & 21.587(23)\\ 
373 & 36 & 5.094(12) & 95.09(33) & 0.27294(44) & 0.07597(14) & 0.2445(20) & 30.390(43)\\ 
512 & 40 & 5.497(12) & 110.66(39) & 0.27303(38) & 0.06615(13) & 0.2101(22) & 41.149(65)\\ 
592 & 42 & 5.675(10) & 118.34(32) & 0.27217(33) & 0.062069(87) & 0.2028(15) & 47.288(60)\\ 
681 & 44 & 5.900(14) & 127.52(47) & 0.27299(37) & 0.05864(11) & 0.1943(19) & 53.826(94)\\ 
778 & 46 & 6.060(21) & 134.56(72) & 0.27292(51) & 0.05503(16) & 0.1776(32) & 61.22(17)\\ 
884 & 48 & 6.176(25) & 140.72(88) & 0.27105(64) & 0.05165(17) & 0.1672(32) & 69.04(20)\\ 
\hline 
  & & \multicolumn{6}{c}{$ \beta = 0.31125 $} \\ 
 \hline 
32 & 16 & 2.4391(21) & 22.020(21) & 0.27017(34) & 0.190759(98) & 0.43198(51) & 1.9496(46)\\ 
46 & 18 & 2.7438(24) & 27.858(25) & 0.27025(36) & 0.168324(83) & 0.40925(54) & 3.1504(59)\\ 
64 & 20 & 3.0473(30) & 34.278(36) & 0.27090(40) & 0.149801(87) & 0.38624(68) & 4.7501(71)\\ 
88 & 22 & 3.3586(35) & 41.500(52) & 0.27182(35) & 0.133692(89) & 0.36231(68) & 6.880(11)\\ 
110 & 24 & 3.6414(40) & 48.512(68) & 0.27334(33) & 0.123605(96) & 0.34876(90) & 8.7956(94)\\ 
168 & 28 & 4.1695(60) & 63.68(13) & 0.27298(39) & 0.10579(11) & 0.3197(11) & 13.791(15)\\ 
262 & 32 & 4.6869(85) & 80.48(23) & 0.27293(37) & 0.08872(13) & 0.2777(13) & 21.586(26)\\ 
373 & 36 & 5.172(11) & 97.82(33) & 0.27350(41) & 0.07711(14) & 0.2516(17) & 30.574(43)\\ 
512 & 40 & 5.604(19) & 115.01(60) & 0.27304(62) & 0.06757(18) & 0.2286(28) & 41.61(13)\\ 
\hline 
  & & \multicolumn{6}{c}{$ \beta = 0.3115 $} \\ 
 \hline 
32 & 16 & 2.4499(21) & 22.206(20) & 0.27029(35) & 0.191660(94) & 0.43570(56) & 1.9444(47)\\ 
46 & 18 & 2.7630(23) & 28.101(26) & 0.27166(33) & 0.169113(85) & 0.41200(53) & 3.1561(62)\\ 
64 & 20 & 3.0710(26) & 34.685(33) & 0.27190(34) & 0.150740(77) & 0.38977(61) & 4.7504(72)\\ 
88 & 22 & 3.3783(39) & 42.010(63) & 0.27167(36) & 0.13457(11) & 0.36717(79) & 6.894(10)\\ 
110 & 24 & 3.6678(42) & 49.176(70) & 0.27356(39) & 0.124528(95) & 0.35446(75) & 8.8076(88)\\ 
168 & 28 & 4.2136(58) & 64.79(12) & 0.27404(38) & 0.10677(10) & 0.32546(96) & 13.840(14)\\ 
262 & 32 & 4.7588(76) & 82.70(20) & 0.27383(39) & 0.09004(12) & 0.2897(13) & 21.687(26)\\ 
373 & 36 & 5.2512(90) & 100.80(27) & 0.27356(32) & 0.07832(11) & 0.2583(15) & 30.613(44)\\ 
512 & 40 & 5.698(13) & 118.80(42) & 0.27327(43) & 0.06871(13) & 0.2344(20) & 41.721(66)\\ 
\hline 
  & & \multicolumn{6}{c}{$ \beta = 0.3118 $} \\ 
 \hline 
32 & 16 & 2.4574(22) & 22.305(18) & 0.27073(38) & 0.192139(82) & 0.43817(50) & 1.9559(55)\\ 
46 & 18 & 2.7764(21) & 28.383(23) & 0.27159(31) & 0.170033(76) & 0.41621(51) & 3.1652(61)\\ 
64 & 20 & 3.0867(27) & 35.021(38) & 0.27207(32) & 0.151519(88) & 0.39259(56) & 4.7673(68)\\ 
88 & 22 & 3.4079(36) & 42.577(56) & 0.27277(35) & 0.135525(97) & 0.37079(73) & 6.9033(91)\\ 
110 & 24 & 3.6859(45) & 49.781(77) & 0.27291(38) & 0.12533(11) & 0.35856(88) & 8.846(11)\\ 
168 & 28 & 4.2503(57) & 66.04(13) & 0.27352(36) & 0.10787(11) & 0.3322(12) & 13.855(15)\\ 
262 & 32 & 4.8231(79) & 84.93(21) & 0.27390(39) & 0.09131(12) & 0.2979(12) & 21.773(29)\\ 
373 & 36 & 5.3518(97) & 104.56(30) & 0.27392(40) & 0.07985(12) & 0.2702(18) & 30.811(53)\\ 
512 & 40 & 5.824(15) & 123.75(52) & 0.27406(42) & 0.07015(16) & 0.2419(23) & 41.981(78)\\ 
\hline 
  & & \multicolumn{6}{c}{$ \beta = 0.312 $} \\ 
 \hline 
32 & 16 & 2.4678(20) & 22.462(19) & 0.27113(34) & 0.192845(85) & 0.43929(51) & 1.9516(55)\\ 
46 & 18 & 2.7878(22) & 28.602(24) & 0.27173(34) & 0.170722(77) & 0.41788(58) & 3.1652(62)\\ 
64 & 20 & 3.1094(26) & 35.444(36) & 0.27279(34) & 0.152516(82) & 0.39803(57) & 4.7690(69)\\ 
88 & 22 & 3.4340(31) & 43.148(50) & 0.27331(33) & 0.136530(86) & 0.37673(70) & 6.9430(85)\\ 
110 & 24 & 3.7134(37) & 50.531(64) & 0.27289(37) & 0.126343(86) & 0.36395(73) & 8.879(11)\\ 
168 & 28 & 4.2984(63) & 67.50(13) & 0.27372(39) & 0.10914(11) & 0.3402(11) & 13.924(14)\\ 
262 & 32 & 4.8875(75) & 87.18(20) & 0.27400(35) & 0.09258(11) & 0.3047(14) & 21.838(24)\\ 
373 & 36 & 5.452(11) & 107.89(34) & 0.27552(39) & 0.08121(14) & 0.2824(19) & 30.973(50)\\ 
512 & 40 & 5.947(17) & 128.62(60) & 0.27501(50) & 0.07164(18) & 0.2575(23) & 42.297(66)\\ 
592 & 42 & 6.211(12) & 140.35(42) & 0.27490(28) & 0.06784(11) & 0.2466(17) & 48.474(63)\\ 
681 & 44 & 6.457(19) & 151.99(71) & 0.27428(46) & 0.06427(16) & 0.2389(26) & 55.32(11)\\ 
778 & 46 & 6.718(37) & 163.8(1.5) & 0.27547(67) & 0.06097(30) & 0.2271(49) & 62.87(24)\\ 
884 & 48 & 6.967(41) & 176.7(1.8) & 0.27476(44) & 0.05773(21) & 0.2143(29) & 70.94(14)\\ 
\hline 
 \end{tabular}
  \caption{The results of our simulations (continued).}
  \label{tab:observables2}
\end{table}

\begin{table}
\begin{center}
\begin{tabular}{cc}
\hline
$L_\bot$   &  $\tau_\chi$   \\
\hline
28 &     2182(82) \\
32 &     3934(190) \\
36 &     6044(359) \\
40 &     9306(738) \\
42 &     11040(615) \\
44 &     12790(986) \\
46 &     21190(3218) \\
48 &     20460(1115) \\
\hline
\end{tabular}
\end{center}
\caption{Integrated autocorrelation times $\tau_\chi$ (expressed in sweeps) 
for the susceptibility for $\beta = 0.312$. }
\label{Table-tauchi}
\end{table}

\begin{figure}[t]
  \centering
\begin{minipage}{\textwidth}
  \epsfig{file=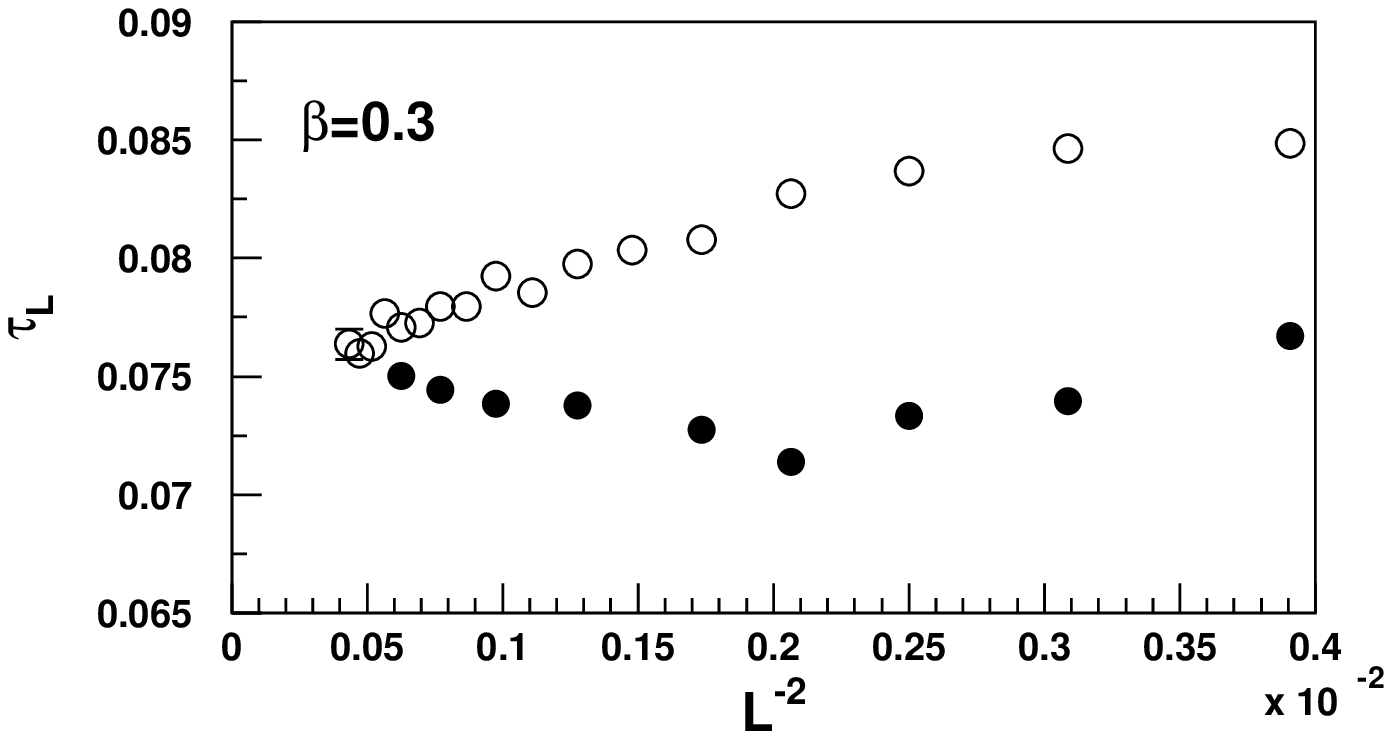,width=0.5\textwidth}
  \epsfig{file=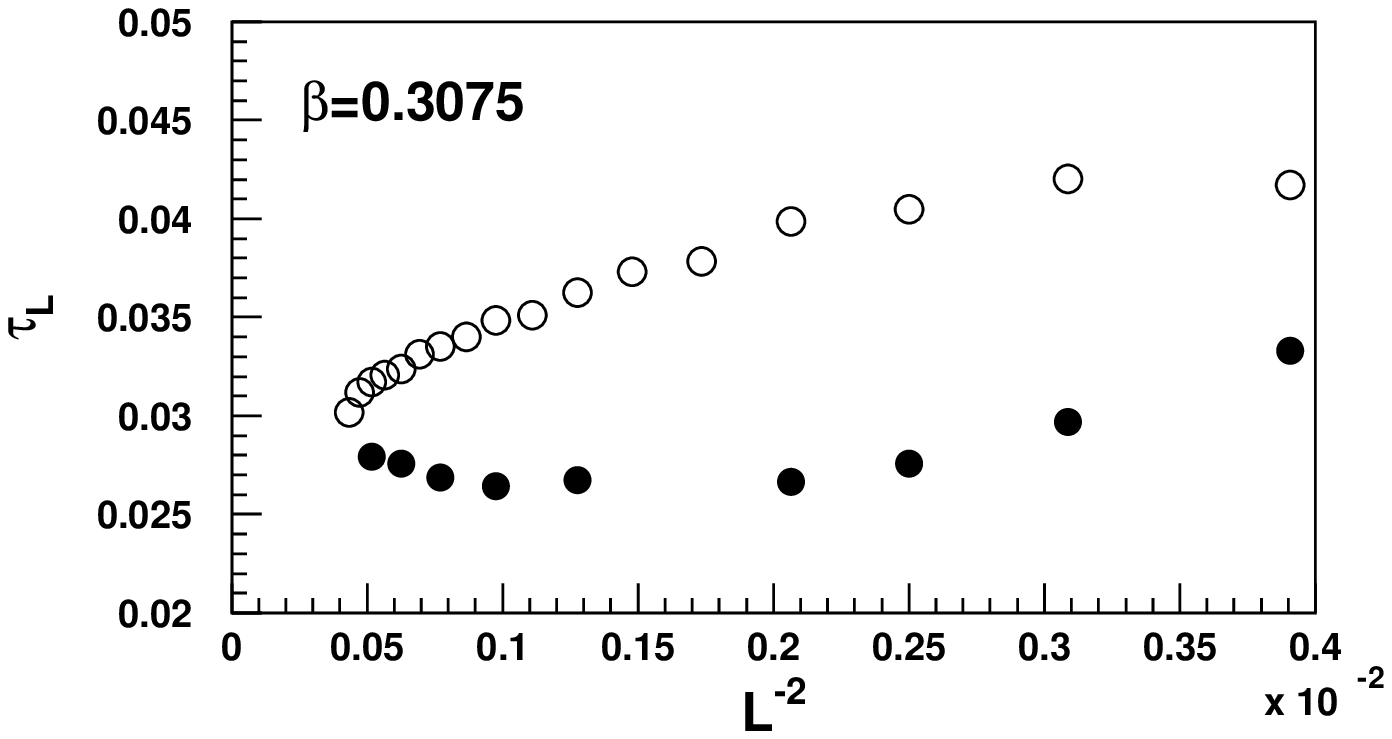,width=0.5\textwidth}
\end{minipage}
\begin{minipage}{\textwidth}
\epsfig{file=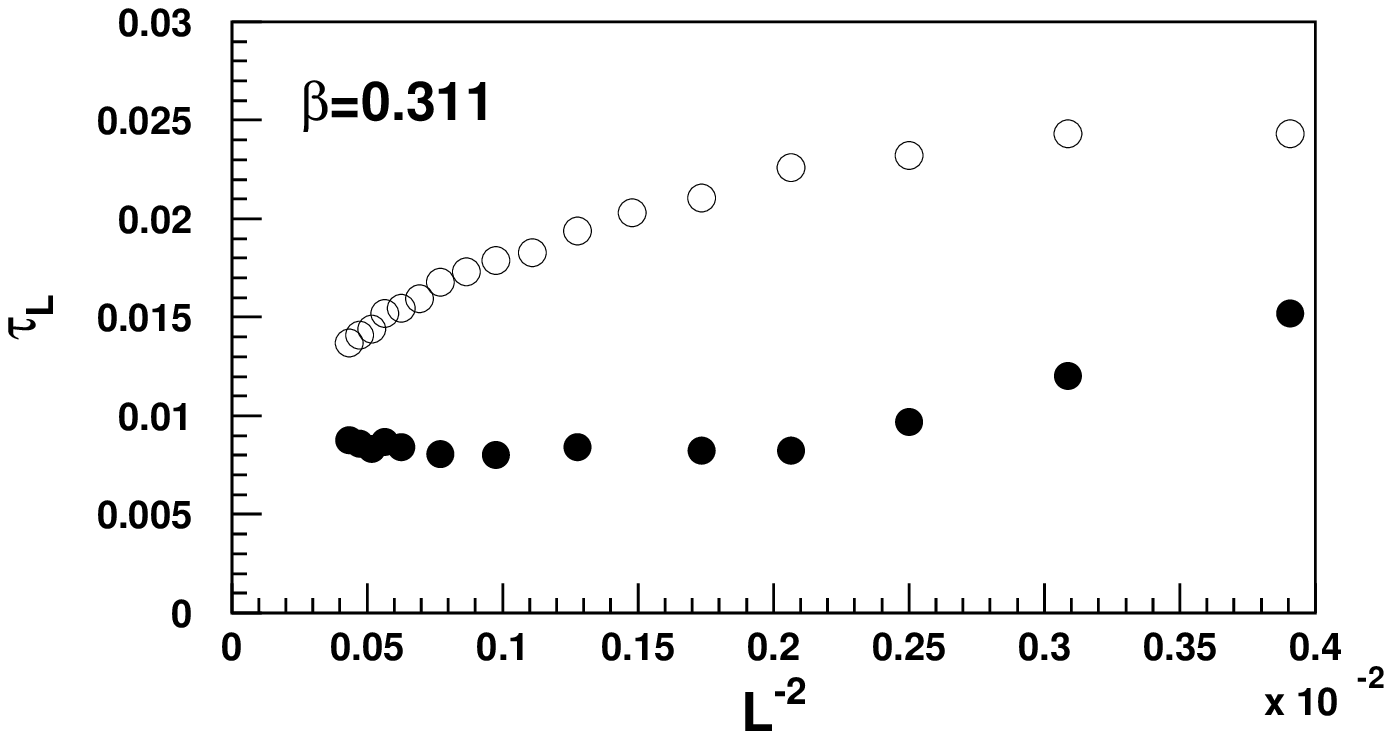,width=0.5\textwidth}
  \epsfig{file=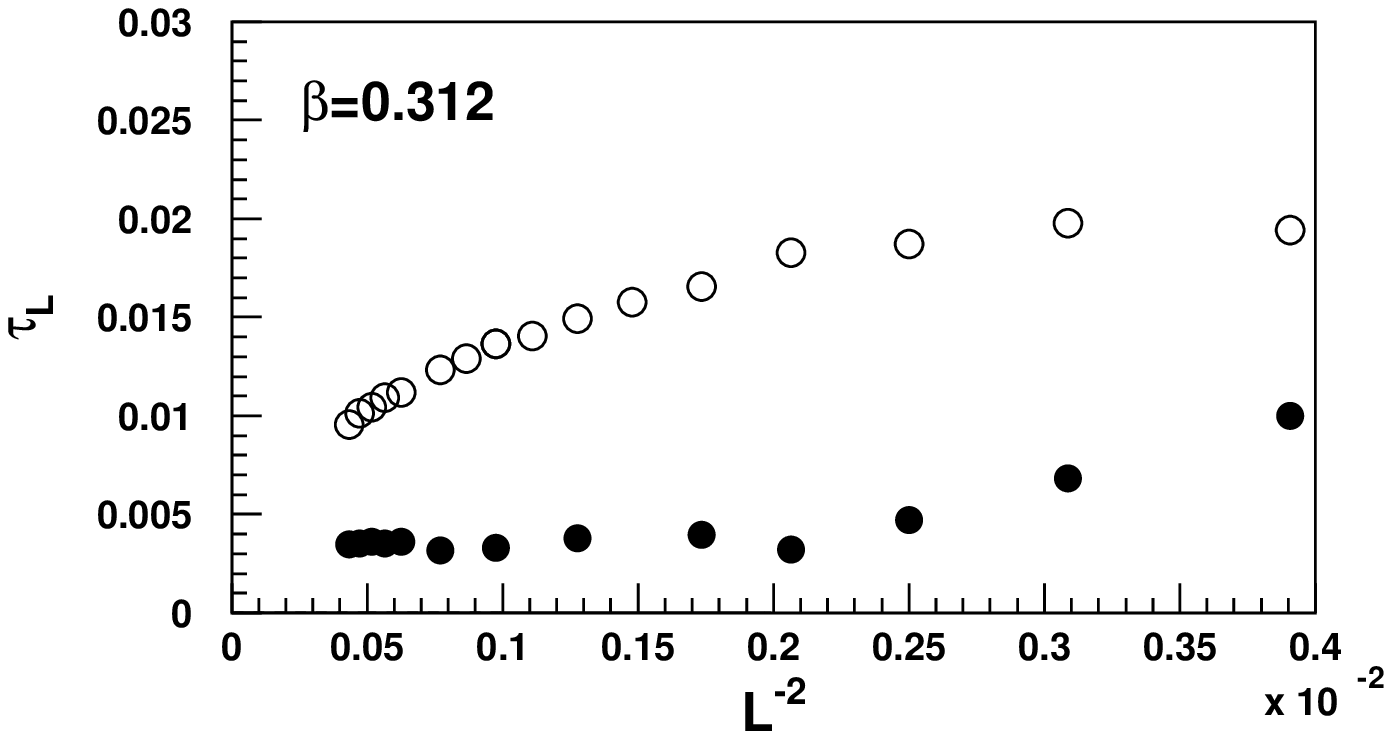,width=0.5\textwidth}
\end{minipage}
\caption{$\tau_L\equiv \tau(\beta;L_\|,L_\bot)$ 
for different  lattices as a function of $L^{-2}\equiv L_\bot^{-2}$.
Filled (respectively empty) points refer to lattices with aspect
ratio $S_{2}$  (respectively $S_{1}$) fixed. Here $S_2 \approx 0.200$,
$S_1 \approx 0.106$. Errors are smaller than the size of the points.
}
\label{fig:tauelle1}
\end{figure}

\begin{figure}[!thb]
  \begin{center}
  \epsfig{file=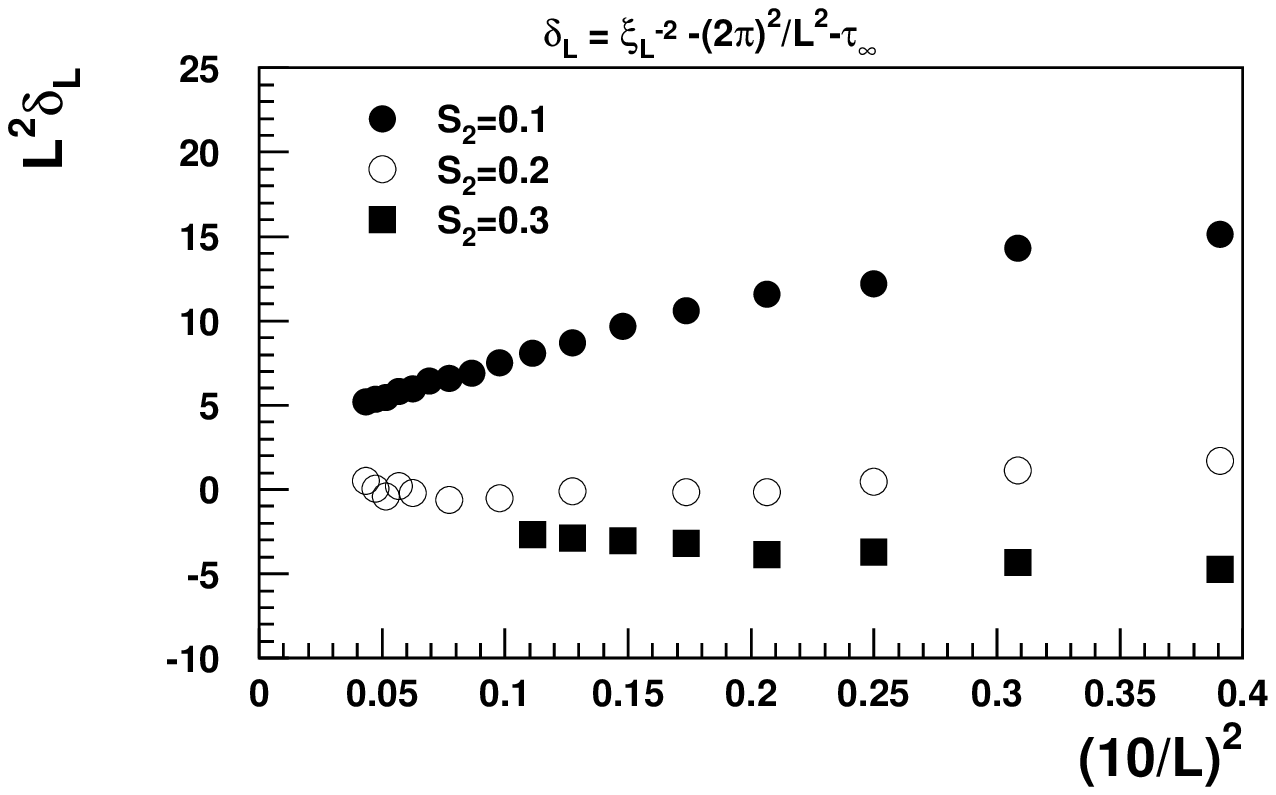,width=0.6\textwidth} 
  \epsfig{file=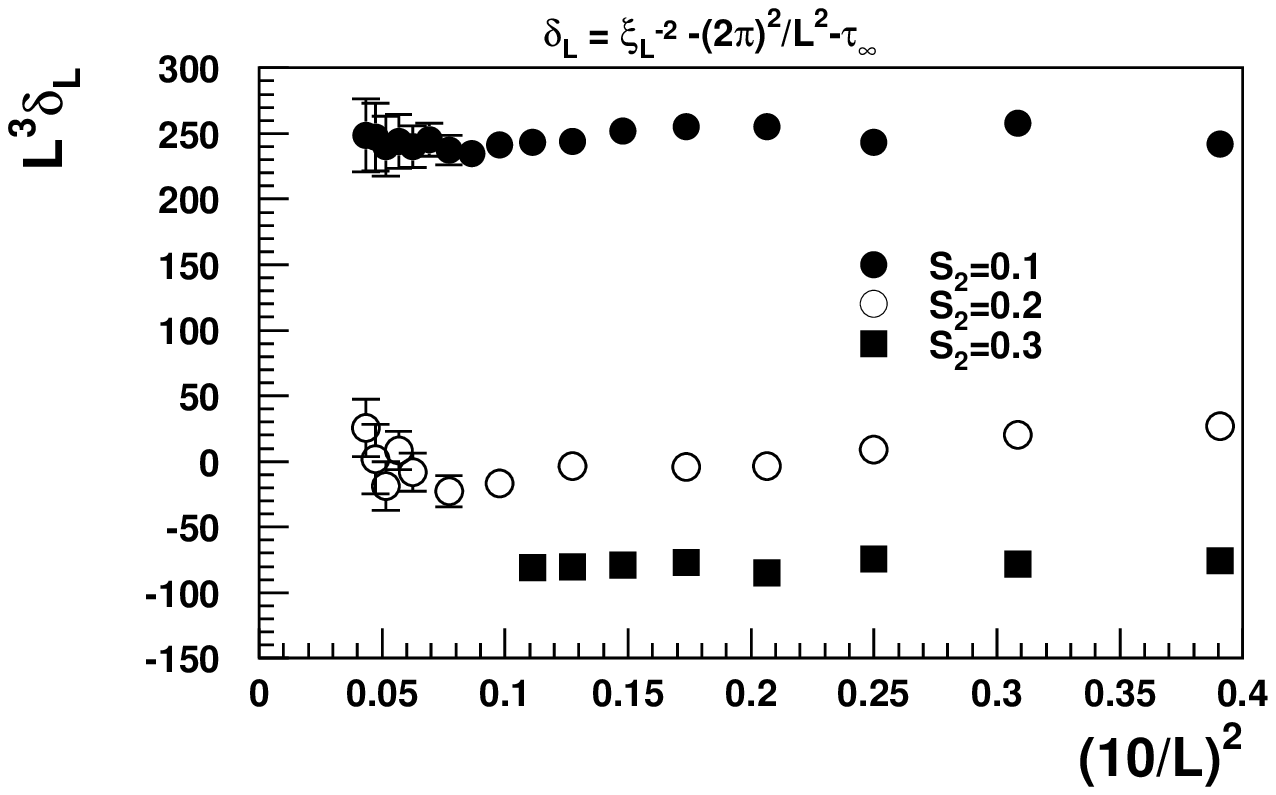,width=0.6\textwidth}
  \epsfig{file=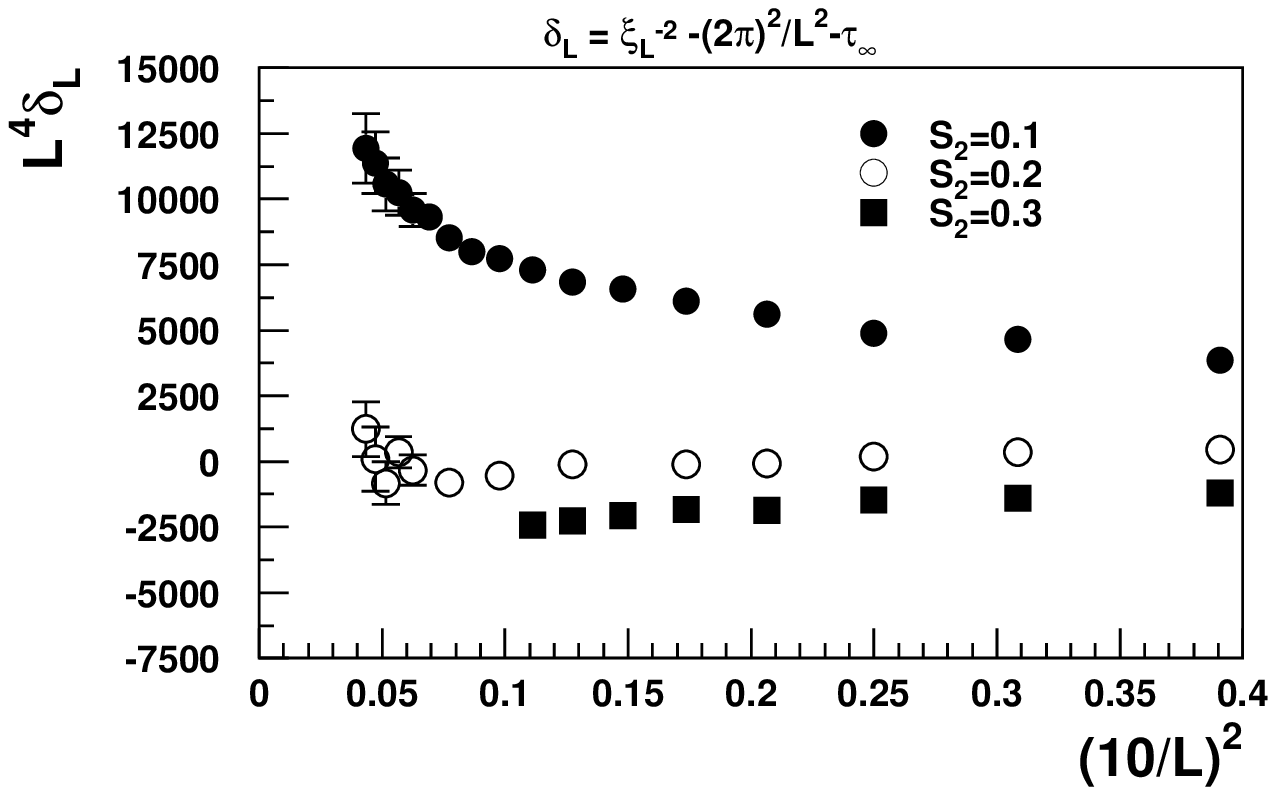,width=0.6\textwidth}
  \end{center}
\caption{$L_\bot^a \delta(\beta;L_\|,L_\bot)$ for different lattices and 
  $S_2$ as a function of
  $L_\bot^{-2}$. Here $a=2,3,4$.}
  \label{fig:various-s2}
\end{figure}
\begin{table}
 \centering
\begin{tabular}[c]{llclllll}
\hline
\hfil$\beta$\hfil & \hfil $L_{\text{min}}$ \hfil & \hfil $ N$ \hfil & \hfil $R^2$ \hfil &  \hfil $R^2_c$ \hfil & \hfil
$\tau_\infty$ \hfil & \hfil $a$\hfil & 
$\xi_{\bot,\infty}(\beta)$\\
\hline
0.28     & 22 & 4 &  3.12  & 9.49  & 0.2397(13)   &    $-$5.44(87) &
       2.043(23) \\
0.29     & 22 & 4 &  2.76  & 9.49  & 0.15309(87)  &    $-$4.19(56) &
       2.556(29) \\
0.3      & 22 & 4 &  5.09  & 9.49  & 0.07607(35)  &    $-$2.02(24) &
       3.626(34) \\
0.3025   & 22 & 4 &  5.19  & 9.49  & 0.05864(36)  &    $-$1.30(25) &
       4.130(51) \\
0.305    & 24 & 8 &  14.63 & 15.50 & 0.04382(26)  &    $-$1.91(26) &
       4.777(57) \\
0.3075   & 26 & 2 &  0.70  & 6.00  & 0.02961(52)  &    $-$3.33(67) &
       5.81(20)  \\
0.31     & 22 & 4 &  8.29  & 9.49  & 0.01309(11)  &    0.310(74) &
       8.74(15) \\
0.3105   & 22 & 4 &  7.57  & 9.49  & 0.01069(18)  &     0.09(13) &
       9.67(33)\\
0.311    & 24 & 7 &  14.78 & 14.08 & 0.00852(21) &    $-$0.17(18) &
        10.83(54)\\
0.31125  & 22 & 4 &  2.65  & 9.49  & 0.00703(11)  &    $-$0.012(72) &
       11.92(39) \\ 
0.3115   & 22 & 4 &  8.56  & 9.49  & 0.00586(14)  &    0.01(13) &
       13.06(63) \\
0.31175  & 24 & 3 &  9.21  & 7.82  & 0.00421(20)  &   0.49(19) &
       15.4(1.5) \\
0.312    & 28 & 6 &  11.46  & 12.59 & 0.00348(16)  &   0.01(20) &
        17.0(1.5) \\
\hline
\end{tabular}
  \caption{Fit of $\tau(\beta;L_\|,L_\bot)$ with 
   $\tau_\infty(\beta) + a(\beta) L_\bot^{-2}$ using the data with
    $S_2\approx 0.200$: $L_{\text{min}}$ is the minimum value of $L$
    used in the fit, $N$ is the number of degrees of freedom, $R^2$
    is the $\chi^2$, i.e. the sum of the square residuals,
    and $R^2_c$ is the value
    of $R^2$ corresponding to the $95\%$ confidence level based on a $\chi^2$
    distribution with $N$ degrees of freedom. In the last column we report
    $\xi_{\bot,\infty}^2(\beta) = 1/\tau_\infty(\beta)$.}
  \label{tab:fit_tauelle}
\end{table}

\begin{table}
\begin{center}
\begin{tabular}{llcll}
\hline
\multicolumn{1}{c}{$\beta$} &
\multicolumn{1}{c}{$\xi_{\bot,\infty}(\beta)$} &
\multicolumn{1}{c}{$N$} &
\multicolumn{1}{c}{$R^2$} &
\multicolumn{1}{c}{$R^2_c$}  \\
\hline
0.28             & 2.0603(39)    & 3 & 1.1 & 7.8 \\
0.29             & 2.5850(50)    & 3 & 1.5 & 7.8 \\
0.3              & 3.6702(52)    & 3 & 3.8 & 7.8 \\
0.3025           & 4.1668(88)    & 3 & 5.6 & 7.8 \\
0.305            & 4.8474(92)    & 8 & 18.1 & 15.5 \\
0.3075           & 6.040(20)     & 3 & 6.4 & 7.8 \\
0.31             & 8.673(34)     & 3 & 2.8 & 7.8 \\
0.3105           & 9.654(54)     & 3 & 3.8 & 7.8 \\
0.311            & 10.866(45)    & 7 & 19.9 & 14.1 \\
0.31125          & 11.97(11)     & 3 & 0.6 & 7.8 \\
0.3115           & 13.07(10)     & 3 & 5.4 & 7.8 \\
0.31175          & 14.81(16)     & 3 & 2.8 & 7.8 \\
0.312            & 17.00(16)     & 7 & 5.3 & 14.1 \\
\hline
\end{tabular} 
\end{center}
\caption{Results of the extrapolation with the iterative method 
of Ref.~\protect\cite{Caracciolo95}
using the theoretical FSS function. Here 
$N$ is the number of points for each $\beta$, $R^2$ the residual, and 
$R^2_{\rm c}$ is the value corresponding to a 95\% confidence level.
}
\label{tab:fss_extrapolation}
\end{table}

\begin{figure}[h!]
  \centering
  \epsfig{file=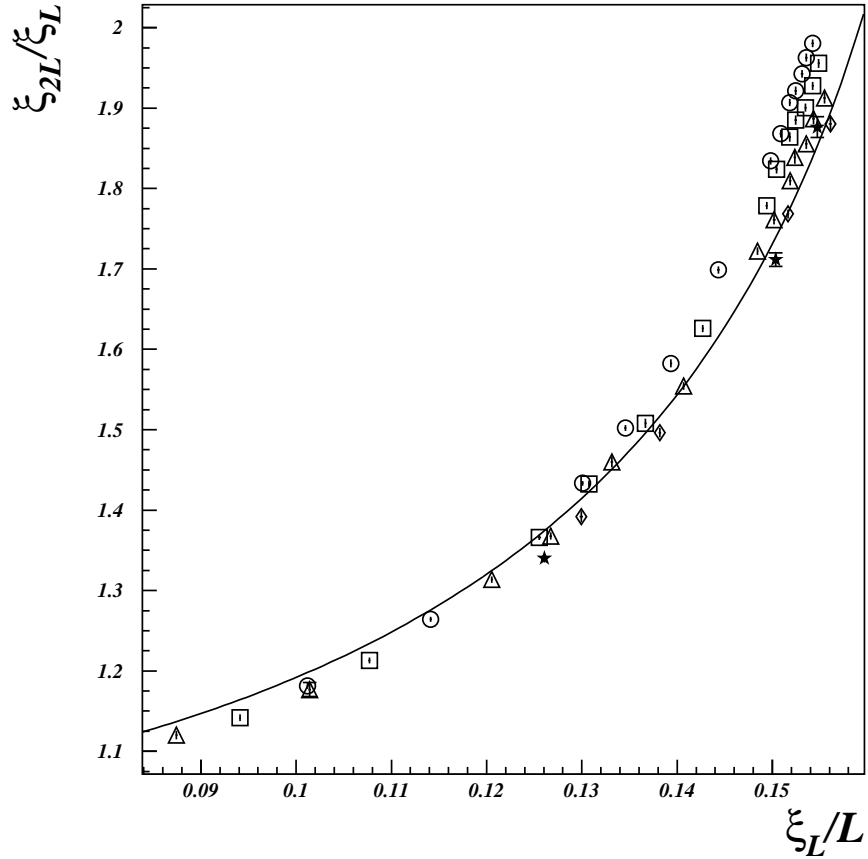,width=0.8\textwidth}
\caption{FSS plot for the transverse correlation length $\xi_\bot$ 
  with $\alpha=2$ and $S_2\approx 0.200$. 
  Here $\xi_L = \xi_\bot(\beta;L_\|,L_\bot)$,
       $\xi_{2L} = \xi_\bot(\beta;8L_\|,2L_\bot)$,
       $L=L_\bot$.
  Different symbols correspond to different lattices sizes:
  $L=$ $16(\circ )$, $18(\square )$, $20(\triangle )$, $22(\lozenge )$,
  $24(\bigstar)$. The
  line is the theoretical prediction $F_\xi(2,\xi_L/L,S_2)$, 
  cf. Eq.~(\ref{eq:theoretical_pred}).}
  \label{fig:newfss2xi}
\end{figure}

\begin{figure}
  \centering
  \epsfig{file=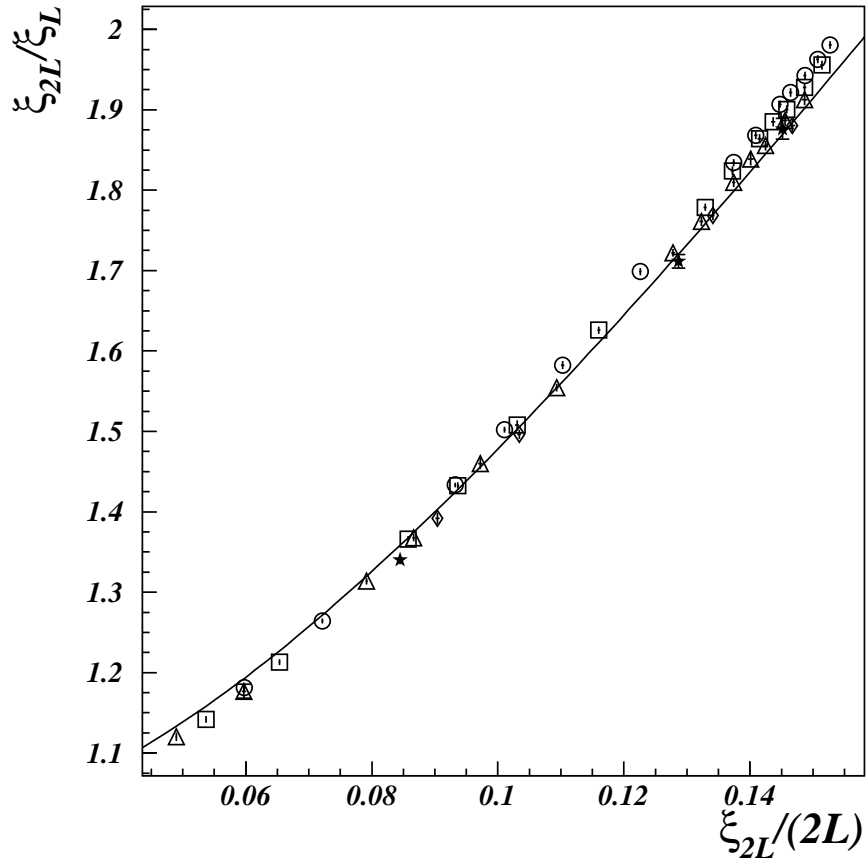,width=0.8\textwidth}
\caption{FSS plot for the transverse correlation length 
  analogous to that presented in 
  Fig.~\protect\ref{fig:newfss2xi}. Here we plot the data vs 
  $\xi_{2L}/(2 L) = \xi(\beta;8L_\|,2L_\bot)/(2 L_\bot)$.
  Symbols are defined as in Fig.~\ref{fig:newfss2xi}. The 
  line is the theoretical prediction }
  \label{fig:newfss2xi2}
\end{figure}

\begin{figure}
  \centering
  \epsfig{file=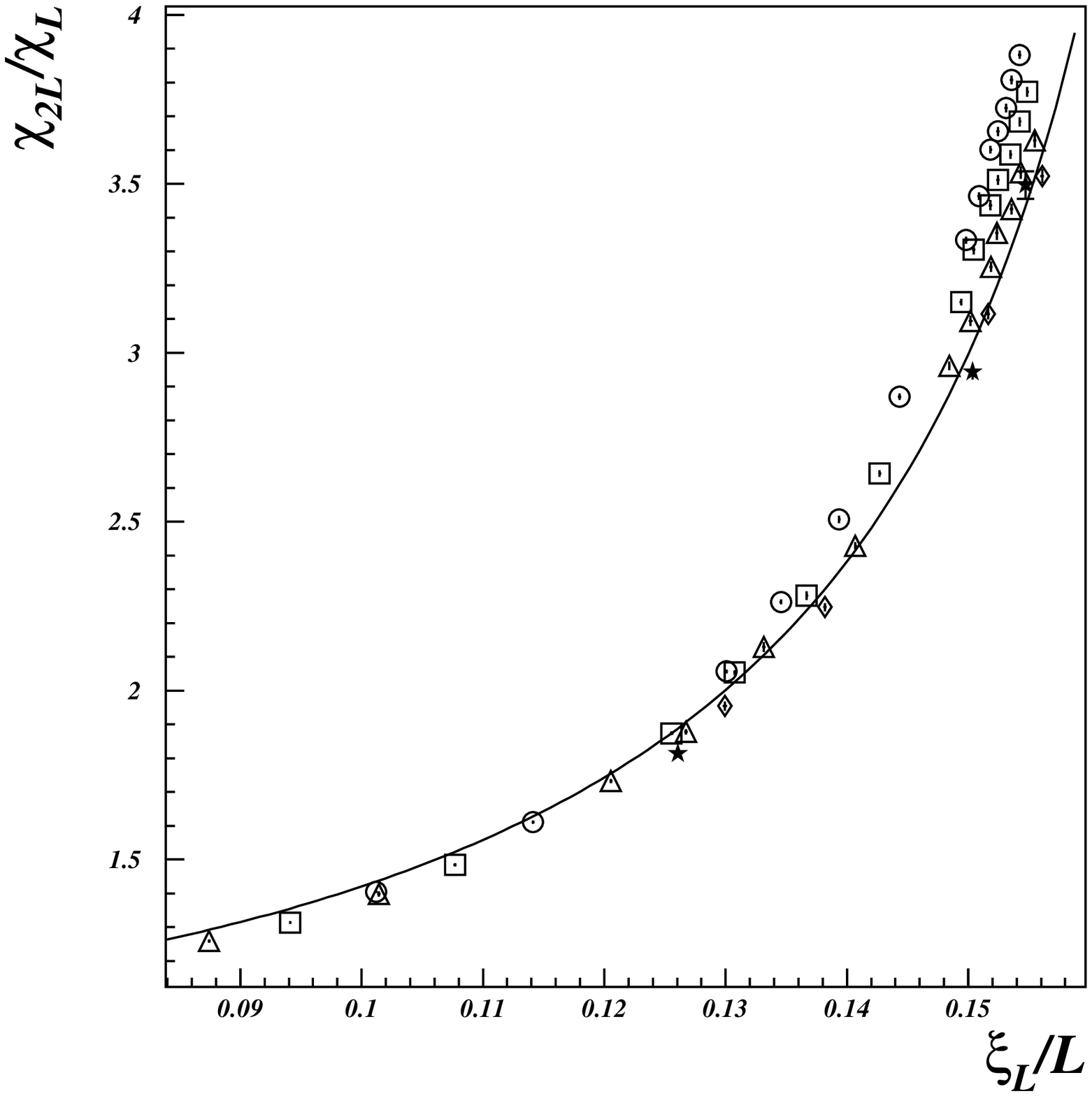,width=0.8\textwidth}
\caption{FSS plot for $\chi$. Symbols are defined as in
  Fig.~\ref{fig:newfss2xi}. The solid
   line is the theoretical prediction, 
   cf.~Eq.~(\ref{eq:theoretical_pred_chi}).}
  \label{fig:newfss2chi2}
\end{figure}

\begin{figure}
  \centering
  \epsfig{file=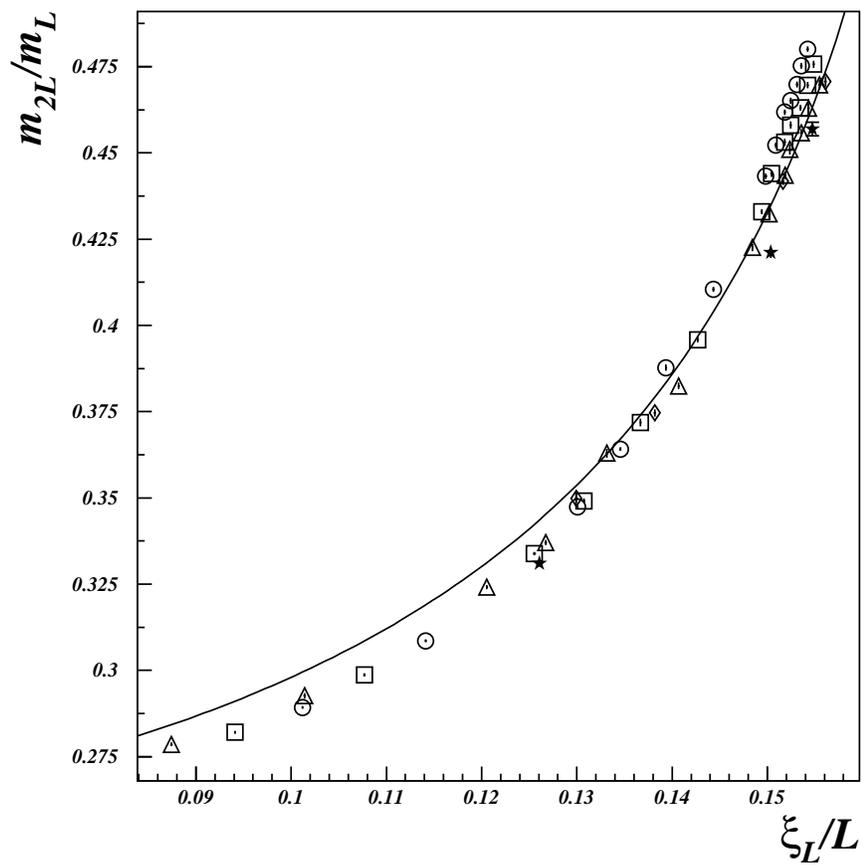,width=0.8\textwidth}
\caption{FSS plot for the magnetization $m$. Symbols are defined as in 
   Fig.~\ref{fig:newfss2xi}.  The solid
   line is the theoretical prediction, 
   cf.~Eq.~(\ref{Fmtheor}).}
  \label{fig:newfss2mag2}
\end{figure}

\begin{figure}
  \centering
  \epsfig{file=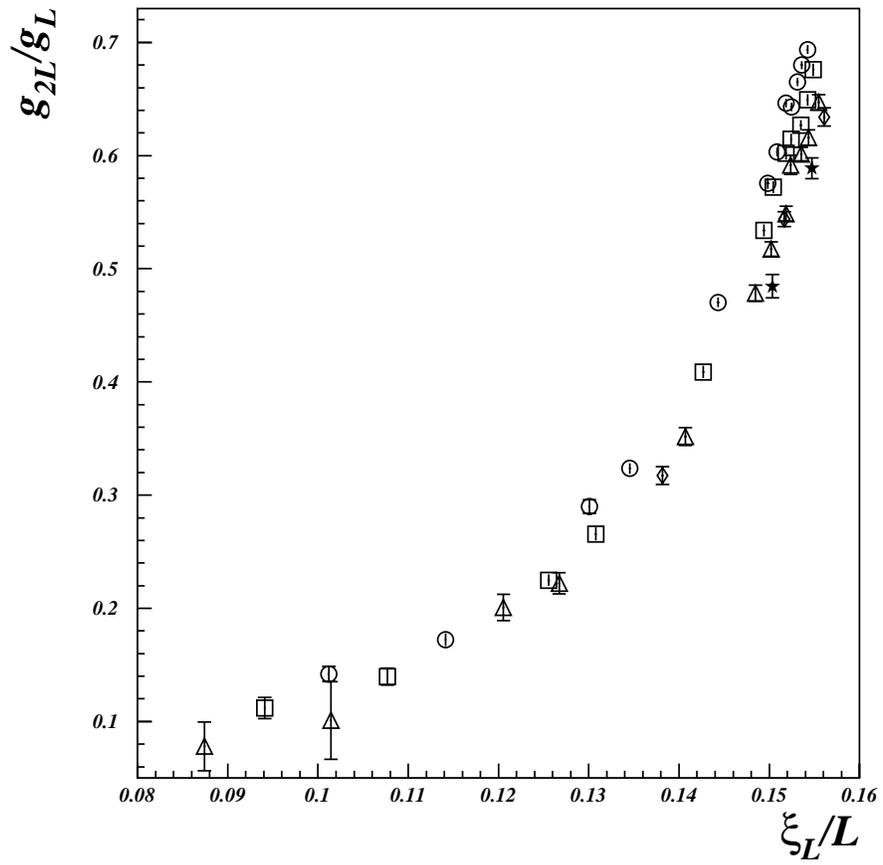,width=0.8\textwidth}
\caption{FSS plot for transverse Binder parameter $g$. Symbols are defined 
   as in Fig.~\ref{fig:newfss2xi}.}
  \label{fig:newfss2binder2}
\end{figure}


\begin{figure}
  \centering
  \epsfig{file=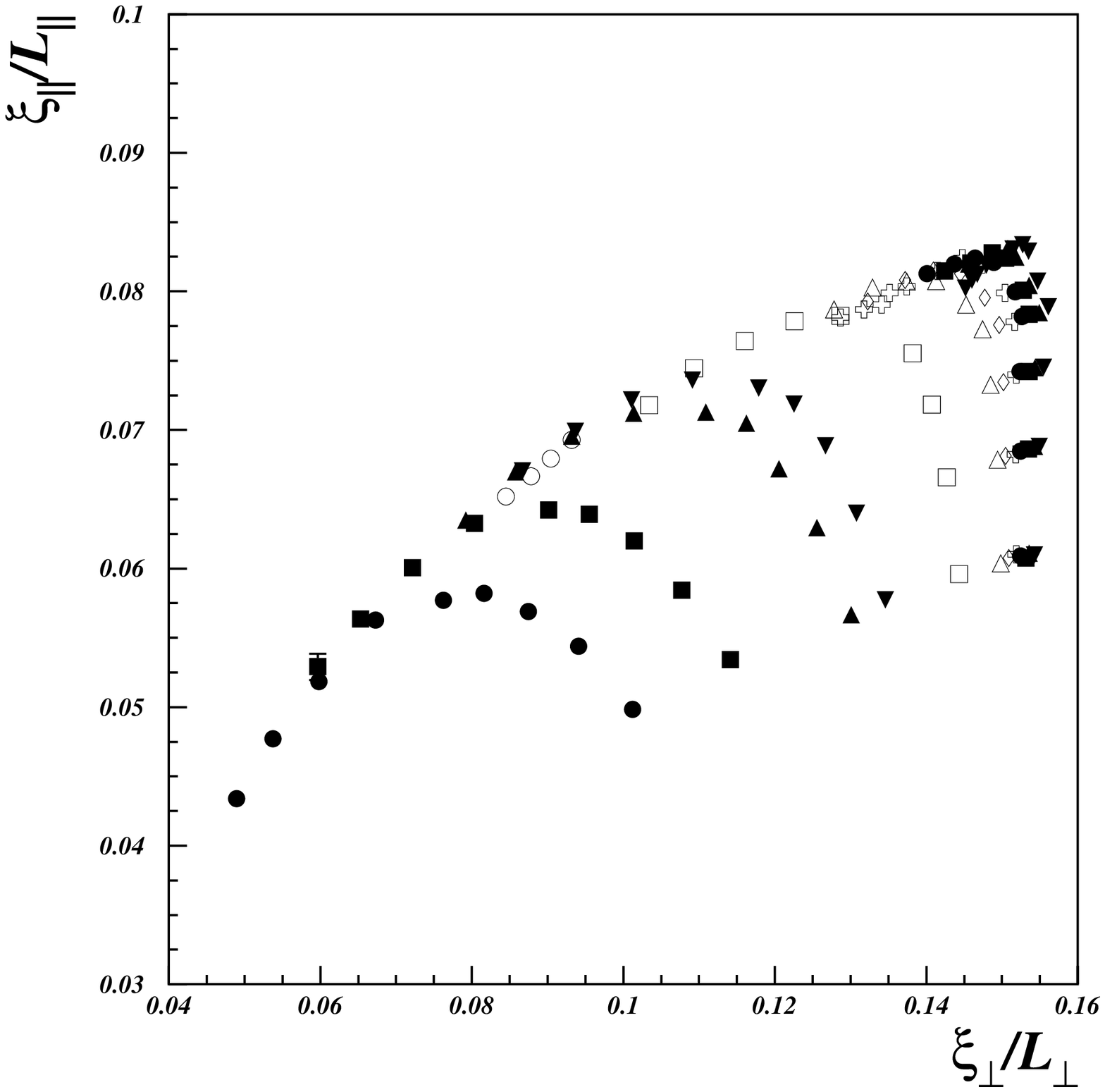,width=0.8\textwidth}
\caption{Plot of $\xi_{\|}(\beta;L_\|,L_\bot)/L_{\|}$ versus 
  $\xi_\bot(\beta;L_\|,L_\bot)/L_\bot$. 
  Each symbol corresponds to a value of $\beta$: 
  0.28 (filled $\bigcirc$), 
  0.29 (filled $\Box$),
  0.30 (filled $\bigtriangleup$), 
  0.3025 (filled $\bigtriangledown$),
  0.305 (empty $\bigcirc$),
  0.3075 (empty $\Box$),
  0.31 (empty $\bigtriangleup$),
  0.3105 (empty $\lozenge$),
  0.311 (empty cross),
  0.31125 (filled $\bigcirc$), 
  0.3115 (filled $\Box$),
  0.31175 (filled $\bigtriangleup$),
  0.312 (filled $\bigtriangledown$).
  For each $\beta$ points move leftwards as 
  $L_\|$ and $L_\bot$ increase.}
  \label{fig:fssdelta}
\end{figure}

\begin{figure}
  \centering
  \epsfig{file=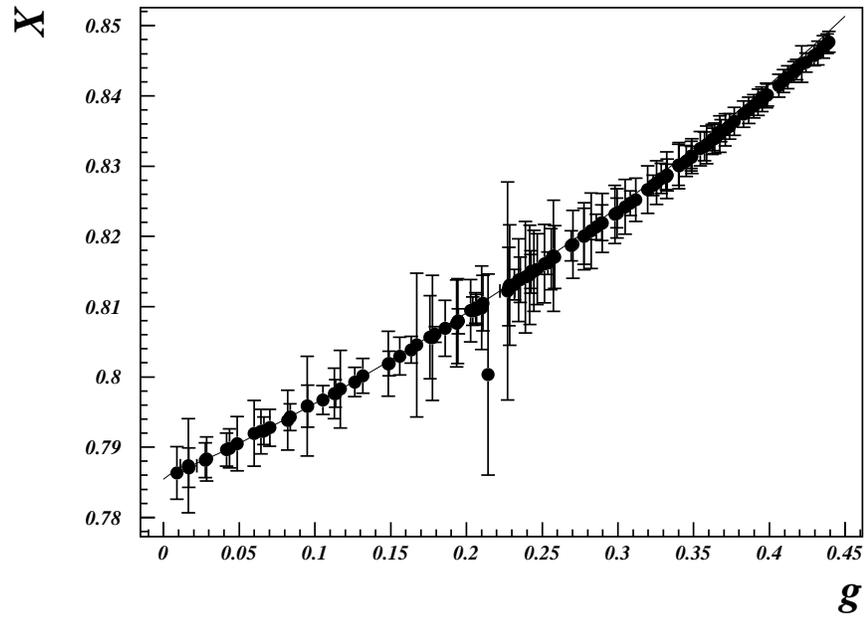,width=0.8\textwidth}
  \epsfig{file=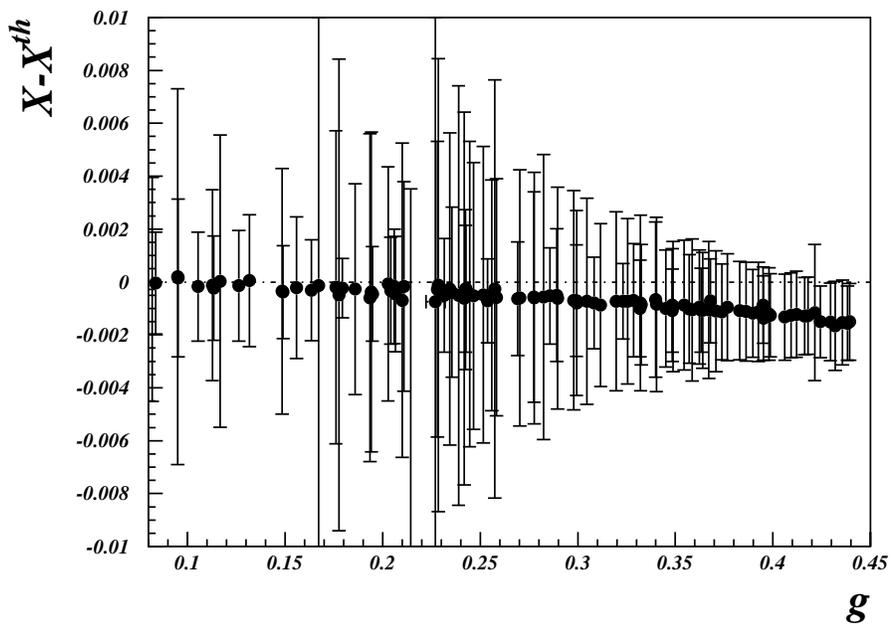,width=0.8\textwidth}
\caption{Upper figure: plot of $X$
  vs. $g$ (see sect.~\ref{sec5:FSS}) together with the 
  theoretical prediction $X^{\text{th}}$, cf.~Eq.~(\ref{gX}).
Lower figure:  plot $X - X^{\text{th}}$ vs. $g$. }
  \label{fig:violets}
\end{figure}



\end{document}